\documentclass[5p,times]{elsarticle}

\usepackage{amsmath,amssymb,amsfonts}
\usepackage{amsthm}
\usepackage{algorithmic}
\usepackage{graphicx}
\usepackage{textcomp}
\usepackage{caption}
\usepackage{subcaption}
\usepackage{etoolbox}
\usepackage{multicol}
\usepackage{url}
\usepackage{tabularray}
\usepackage[linesnumbered,ruled,vlined]{algorithm2e}
\usepackage{comment}
\usepackage[colorlinks=true, linkcolor=black, citecolor=blue, urlcolor=blue]{hyperref} 
\usepackage{nomencl}
\usepackage{framed}
\usepackage{booktabs}
\usepackage{array}
\usepackage[section]{placeins}

\usepackage[most]{tcolorbox}
\usepackage{xcolor}
\tcbset{
  llmchatbox/.style={
    enhanced,
    colback=white,
    colframe=black,
    fonttitle=\bfseries,
    boxrule=0.5pt,
    arc=2mm,
    width=\textwidth,
    before skip=10pt,
    after skip=10pt,
    sharp corners=south,
    coltitle=black,
  },
}

\makenomenclature

\journal{Sustainable Energy Technologies and Assessments}

\begin{document}

\title{Large Language Model-Assisted Planning of Electric Vehicle Charging Infrastructure with Real-World Case Study}

\author[aff1]{Xinda Zheng \fnref{coauthor1}}
\author[aff1]{Canchen Jiang \fnref{coauthor1}}
\author[aff1,aff2]{Hao Wang\corref{cor1}}
\ead{hao.wang2@monash.edu}

\affiliation[aff1]{organization={Department of Data Science and AI, Faculty of Information Technology, Monash University},
    city={Clayton},
    postcode={3800}, 
    state={Victoria},
    country={Australia}}
    
\affiliation[aff2]{organization={Monash Energy Institute, Monash University},
    city={Clayton},
    postcode={3800}, 
    state={Victoria},
    country={Australia}}

\fntext[coauthor1]{These authors contributed equally to this work.}
\cortext[cor1]{Corresponding author: Hao Wang.}

\begin{abstract}
The growing demand for electric vehicle (EV) charging infrastructure presents significant planning challenges, requiring efficient strategies for investment and operation to deliver cost-effective charging services. However, the potential benefits of EV charging assignment, particularly in response to varying spatial-temporal patterns of charging demand, remain under-explored in infrastructure planning. This paper proposes an integrated approach that jointly optimizes investment decisions and charging assignments while accounting for spatial-temporal demand dynamics and their interdependencies. To support efficient model development, we leverage a large language model (LLM) to assist in generating and refining the mathematical formulation from structured natural-language descriptions, significantly reducing the modeling burden. The resulting optimization model enables optimal joint decision-making for investment and operation. Additionally, we propose a distributed optimization algorithm based on the Alternating Direction Method of Multipliers (ADMM) to address computational complexity in high-dimensional scenarios, which can be executed on standard computing platforms. We validate our approach through a case study using 1.5 million real-world travel records from Chengdu, China, demonstrating a 30\% reduction in total cost compared to a baseline without EV assignment.

\end{abstract}


\maketitle

\setlength{\nomitemsep}{0.01cm}
\renewcommand\nomgroup[1]{%
  \item[\bfseries
  \ifstrequal{#1}{I}{Indices}{%
  \ifstrequal{#1}{P}{Parameters}{%
  \ifstrequal{#1}{S}{Sets}{%
  \ifstrequal{#1}{V}{Variables}{}}}}%
]}

\nomenclature[I]{\(i\)}{Index of locations}
\nomenclature[I]{\(j\)}{Index of locations}
\nomenclature[I]{\(t\)}{Index of time slots}
\nomenclature[I]{\(k\)}{Index of interactions for distributed algorithm}
\nomenclature[V1]{\(c_i\)}{Charging capacity decision at location $i$}
\nomenclature[V2]{\(\tilde{c_i}\)}{Auxiliary variable for $c_i$}
\nomenclature[V3]{\(\lambda_i\)}{Lagrangian multipliers (dual variable) for constraint $c_i - \tilde{c_i} = 0$}
\nomenclature[V4]{\(z^t_{ij}\)}{EV assignment decision from location $i$ to location $j$ at time $t$}
\nomenclature[V5]{\(Q_{primal}\)}{Primal error for distributed ADMM algorithm}
\nomenclature[V5]{\(Q_{dual}\)}{Dual error for distributed ADMM algorithm}
\nomenclature[P]{\(\pi_0\)}{Based investment cost per unit charging capacity}
\nomenclature[P]{\(\pi_i\)}{Location-dependent investment cost per unit charging capacity of location $i$}
\nomenclature[P]{\(L_{ij}\)}{Assignment cost per EV location $i$ and location $j$}
\nomenclature[P]{\(\alpha^t_i\)}{Coefficient between charging demand and traffic flow at time $t$ in location $i$}
\nomenclature[P]{\(\beta\)}{Coefficient between charging demand and charging capacity}
\nomenclature[P]{\(F^t_i\)}{Traffic flow at time $t$ in location $i$}
\nomenclature[P]{\(B\)}{Total investment budget}
\nomenclature[P]{\(R\)}{Assignment range limitation} 
\nomenclature[P]{\(C_i\)}{Maximum charging capacity at location $i$}
\nomenclature[P]{\(E_t\)}{Expect number of recurrent periods for time slot $t$}
\nomenclature[P]{\(\rho\)}{Penalty for distributed ADMM algorithm}
\nomenclature[S]{\(\mathcal{T}\)}{Time slots in the operation horizon}
\nomenclature[S]{\(\mathcal{I}\)}{Locations}
\printnomenclature

\section{Introduction}\label{sec:Introduction}
Electric vehicles (EVs) have emerged as a promising solution to reduce greenhouse gas emissions in the transportation sector, thereby advancing the goal of net-zero transportation. The adoption of EVs can reduce operational and maintenance costs for consumers but offer significant environmental benefits for society \cite{xie_how_2024}. Research indicates that widespread EV adoption could mitigate up to 25.5\% of anticipated winter warming \cite{Chen_Yang_2022}. Leading this transformative change, countries such as the United Kingdom are actively transitioning away from fossil fuels toward an electrified transportation future \cite{noauthor_reducing_2018}. However, the success of this transition is heavily dependent on the development of robust EV charging infrastructure. Proactive and strategic deployment of charging stations is indispensable for enabling a smooth shift to electric mobility \cite{ou_modeling_2020}.

Despite recent progress, including a notable 55\% increase in public charging stations reported by the International Energy Agency’s Global EV Outlook 2023 \cite{noauthor_global_nodate}, significant challenges persist. These include determining optimal locations for charging stations, managing service coverage and efficiency, and ensuring the economic viability of charging infrastructure investments \cite{shafiei_fast-charging_2022}. According to \cite{jiang2024urban}, residents’ satisfaction with the availability of charging stations significantly influences their willingness to adopt EVs. Leading EV manufacturers have advocated for increased government incentives and substantial investments in charging infrastructure to boost EV adoption rates worldwide \cite{european_carmakers}. Conversely, recent reports \cite{electrek2024, pew2024, evmagazine2024} suggest that the rapid expansion of public EV charging infrastructure across different countries has led to overcapacity, resulting in inefficiencies and underutilized charging stations. These misalignments with actual EV user demand have caused resource waste and run counter to sustainability. Therefore, thorough planning and deployment of EV charging infrastructure are crucial for providing quality charging services and promoting a sustainable and efficient charging ecosystem. Such an ecosystem not only supports the growing EV market but also ensures that the environmental and social benefits of EVs are fully realized, ultimately contributing to global efforts to sustainable development.

Numerous optimization models have been proposed to address these issues. Early work applied flow-based siting to locate facilities at centroids of traffic demand \cite{dantrakul_applied_2014}, while operational studies optimized EV assignment and charging for fixed stations using equilibrium formulations \cite{chen_optimal_2016}. More recent efforts attempted to link investment and operations, such as facility-level joint designs \cite{nishimwe_h_combined_2021}, joint optimization of autonomous EV fleets with station siting \cite{luke_joint_2021}, and bi-level formulations integrating location and routing \cite{LI2022104153}. These advances (see Section~\ref{Related work} for a full review) nevertheless fall short in two critical respects. First, most inadequately capture the spatial-temporal coupling of charging, even though real-world charging demand varies strongly across both time and space. Second, building such detailed models typically requires extensive expert effort to define problem structures, extract relevant features, and handcraft optimization formulations. These limitations hinder efficient and scalable planning at city scale.

To overcome this modeling bottleneck, recent advances in large language models (LLMs) provide promising tools to accelerate formulation. Several studies  \cite{ahmed_lm4opt_2024, spreitzer_mathematical_2024, huang-etal-2025-llms} have demonstrated that LLMs are capable of translating structured natural-language descriptions into optimization variables, constraints, and solver-ready code. Building on this capability, we propose a human-LLM collaborative workflow for EV charging infrastructure planning. This approach reduces the burden of manual formulation while allowing experts to iteratively refine and validate model structures, enabling rapid development of context-specific optimization models. However, even with faster model generation, realistic planning problems that account for fine-grained spatial and temporal demand remain very large-scale and high-dimensional, creating significant computational challenges.

Addressing this computational complexity requires solution methods that scale beyond centralized solvers. Distributed optimization, particularly the Alternating Direction Method of Multipliers (ADMM) \cite{Boyd_Parikh_Chu_Peleato_Eckstein_2010}, is well suited to such settings. ADMM decomposes large models into parallelizable subproblems, reducing computational burdens while maintaining theoretical guarantees. It has already proven effective in diverse energy system applications, including microgrid energy trading \cite{wang2016incentivizing}, restoration in transmission systems with high renewable penetration \cite{WANG2024109532}, energy management for virtual power plants \cite{yang2021blockchain}, and air-conditioning system optimization \cite{SONG2022119831}. By extending these advantages to EV infrastructure planning, we enable efficient solution of city-scale problems where spatial-temporal coupling across many locations and times would otherwise render optimization intractable.

Motivated by these considerations, this paper develops a novel and practical approach that integrates LLM-assisted modeling with distributed optimization to jointly address infrastructure investment and EV charging assignment. Our model, developed through an LLM-assisted modeling process, explicitly captures spatial-temporal variations in charging demand and the coupling between investment and operational decisions across time and space, improving both modeling efficiency and accuracy. To address the computational complexity of such high-dimensional problems, we design a distributed optimization algorithm based on the ADMM, which enables efficient decomposition into regional sub-problems solvable on standard computing hardware. The effectiveness of our framework is demonstrated through a real-world case study in Chengdu, China, using 1.5 million travel records, showing the substantial benefits of jointly optimizing EV infrastructure planning and operational assignment. The main contributions of this paper are as follows:

\begin{itemize}
\item \emph{LLM-assisted joint optimization of charging infrastructure investment and EV assignment}: We propose a comprehensive optimization framework that jointly captures long-term infrastructure investment and short-term EV assignment, grounded in a nuanced understanding of spatial-temporal charging demand. This formulation explicitly models the interdependencies between investment and operation across multiple time scales and locations. To support efficient model development, we leverage a LLM to generate and iteratively refine the mathematical formulation based on natural-language descriptions of the planning scenario. The resulting model integrates expert-driven insights with LLM-assisted structure generation, enabling a rigorous and scalable approach to EV infrastructure planning.

\item \emph{An ADMM-based distributed solution method}: To overcome the computational burdens of optimizing a large-scale optimization problem with temporal-spatial coupling, we design a distributed algorithm based on ADMM. This algorithm mitigates computational overhead by decomposing the central problem into sub-problems that can be solved in parallel at individual nodes. This approach ensures that each sub-problem remains manageable within the computational capabilities of standard computing systems, allowing for broader and more feasible implementation.

\item \emph{A real-world case study in Chengdu, China}: Using an empirical dataset of 1.5 million real-world travel records from Chengdu, China, we demonstrate the effectiveness of our method in an urban setting. Our study shows that incorporating EV assignments results in a 30\% reduction in overall costs compared to a baseline that overlooks the operational benefits. This joint optimization not only reduces costs but also improves charging facility utilization. The results analysis provide insights into spatial-temporal characteristics, offering a deeper understanding of charging demand. Additionally, the case study showcases the effectiveness of our ADMM-based algorithm in a real-world context.
\end{itemize}

The remainder of this paper is organized as follows. Section~\ref{Related work} provides a review of relevant literature on EV infrastructure planning. Section~\ref{Optimization} presents the LLM-assisted modeling for the joint investment and assignment optimization problem. Section \ref{ADMM} is dedicated to the exposition of the distributed ADMM algorithm for solving the formulated optimization problem. In Section~\ref{Case study}, we apply the proposed model to a case study, utilizing real-world travel records from Chengdu, China, and solving an optimal solution for the given scenario. Lastly, Section~\ref{Conclusion} concludes this paper with outlooks for future work.

\section{Literature review}\label{Related work}
\subsection{Siting and demand estimation approaches}
Early efforts in EV charging infrastructure planning focused on estimating demand and siting facilities based on transportation and demographic data. Flow-based models placed charging facilities at demand centroids to capture traffic coverage \cite{dantrakul_applied_2014}, while OD-matrix formulations extended this idea to optimize station placement based on traffic flows \cite{krishnakumari_data_2019}. Further OD-based optimization approaches, such as \cite{jing_location_2017} and \cite{zhang2023optimal}, share the similar ideas which was used to maximize service coverage along major routes. Alternative methods adopted Geographic Information Systems (GIS) and population-based approaches to incorporate sustainability and user considerations. For example, \cite{charly2023identifying} proposed a two-stage GIS method for sustainable site selection, while \cite{GUO2023104962} introduced an anticipatory framework exploiting charging-while-parking patterns to balance access and productivity. These approaches provide valuable spatial insights, but their reliance on static demand inputs means that temporal fluctuations and operational assignment are largely overlooked.

\subsection{Joint planning, operational assignment, and temporal dynamics}
To better capture uncertainty and dynamic conditions, a variety of optimization frameworks have been developed. Stochastic models and clustering approaches have been used to co-plan charging and renewable integration \cite{ASAAD2024112908}, microgrids with storage and prosumers \cite{HABIB2023107962}, and long-term distributed resources with demand response \cite{BASWAIMI2024113562}. Bi-level formulations link station siting with coupled distribution-transportation networks \cite{LI2024110442} or private investment profitability \cite{MAHDIPOUR2024110309}, while other studies apply multi-objective optimization \cite{YI2024110940}, reinforcement learning \cite{HEO2024105567}, or hybrid deep learning-queuing \cite{POURVAZIRI2024103568}methods to trade off reliability, economy, and waiting times.  

Beyond siting, several works emphasize the behavioral and system-level aspects of charging. Studies consider driver attitudes in route choice \cite{WANG2024186}, rolling-horizon coordination for uncertain arrivals \cite{jiang2025dynamic}, and demand response to improve resiliency \cite{ELKADEEM2023108198}. Broader frameworks address battery swapping co-planning with grids \cite{SHAKER2023106455} and joint optimization of fleet size, capacity, and operations \cite{paparella_joint_2022}, while \cite{PATIL2023104265} warns that fragmented supply- and demand-side approaches risk misalignment without integrated system planning.

Some research directly integrates EV assignment into planning. Equilibrium \cite{wei_network_2018} and agent-based \cite{chaudhari_agent-based_2019} models capture interactions between transport and power networks or individual responses. Other contributions adopt macroscopic flow assignment \cite{yang_joint_2021}, stochastic traffic assignment \cite{wang_stochastic_2018}, or bi-level formulations \cite{Li_Liu_Wang_2022, ferro_bi_2020} linking routing, charging, siting, as well as electrical networks \cite{ferro_optimal_2022} and ride-hailing deployment \cite{LAI2024102928}. While these studies highlight the benefits of joint investment-assignment, they often aggregate demand annually, simplify temporal coupling, or suffer from scalability challenges.  

Temporal frameworks further recognize that infrastructure investment and operations occur on distinct horizons. Multi-period models, such as \cite{chung_multi-period_2015} and \cite{ZHANG2023108565}, capture incremental expansion and per-period assignments \cite{li_multi-period_2016}, while microgrid \cite{wang_joint_2017} and EV fleet \cite{luke_joint_2021} studies emphasize joint benefits across horizons. However, reliance on fictional or small-scale data \cite{cheng_electric_2022} and discrete scenario sampling cumulative \cite{LIU2022104012}, spatial-temporal demand remain largely unexplored.

\subsection{Methodological enablers: LLM-assisted modeling and distributed optimization}

While most existing EV planning studies rely on expert-driven model construction, recent advances in LLMs suggest a new paradigm for optimization modeling. GPT-4 has demonstrated the ability to translate structured natural-language problem descriptions into solver-ready formulations \cite{huang_large_2023, ramamonjison2023}, and benchmark studies confirm the effectiveness of prompt-based approaches \cite{ahmed_lm4opt_2024, huang-etal-2025-llms}. Study \cite{liu_mm-agent_2025} further decompose the modeling process into stages of analysis, construction, solving, and reporting, while interactive studies \cite{collins_evaluating_2024, spreitzer_mathematical_2024} emphasize the importance of human-LLM collaboration for correctness and efficiency. These findings consistently indicate that LLMs can accelerate formulation, but also reveal limitations: without expert guidance, generated models often lack rigor or contextual fidelity. Our work responds to this gap by proposing a human-LLM co-development workflow explicitly tailored to EV infrastructure planning, where domain knowledge and iterative refinement remain indispensable. Importantly, the expert remains responsible for validation and refinement, ensuring that the resulting models are both rigorous and contextually appropriate.

On the computational side, prior studies confirm that large-scale spatial-temporal optimization often exceeds the capacity of centralized solvers. The ADMM has proven effective in decomposing high-dimensional problems into parallelizable subproblems, with theoretical convergence guarantees in convex settings \cite{Boyd_Parikh_Chu_Peleato_Eckstein_2010, shi2014linear}. It has been successfully applied in energy systems for decentralized power systems optimization \cite{wang2017fully, sun2021two}, achieved global convergence guarantees \cite{hong2016}, and extended to handle nonconvex formulations \cite{wang2019}. It is also demonstrated in various applications, such as microgrid energy trading \cite{wang2016incentivizing}, transmission restoration with high renewable penetration \cite{WANG2024109532}, virtual power plant management \cite{yang2021blockchain}, and HVAC optimization \cite{SONG2022119831}. Compared with heuristic or deep learning-based approaches, which often lack convergence guarantees, require problem-specific tuning, or depend on large datasets with limited transferability, ADMM offers three key advantages for EV infrastructure planning. Firstly, it exploits the separable structure of joint investment-assignment formulations to achieve scalable parallel computation. Additionally, it provides interpretable optimization outputs directly usable for planning decisions. Furthermore, it ensures robustness and convergence properties not available in heuristic or learning-based methods. Despite these advantages, ADMM has rarely been applied to EV charging infrastructure planning with explicit spatial-temporal assignment. Our work closes this gap by tailoring ADMM to decompose joint investment-assignment problems, enabling tractable and scalable solutions at the city scale.

\subsection{Summary}
The literature establishes strong foundations in siting, behavioral modeling, and partial joint formulations, yet three gaps persist that materially affect practice. First, investment and operational assignments are seldom co-optimized at fine spatial-temporal resolution; as a result, utilization and cost outcomes remain sensitive to unmodeled spatial-temporal demand patterns. Second, constructing high-fidelity models is still expert-intensive, and existing uses of LLMs have not been systematized within EV planning workflows to reduce formulation burden while preserving reliability. Third, city-scale joint models with explicit spatial-temporal coupling remain computationally prohibitive for centralized solvers, and decomposition methods are rarely tailored to investment-assignment structure.

Accordingly, we develop an LLM-assisted joint optimization framework that explicitly couples long-horizon investment with short-horizon EV assignment over space and time. Building on this framework, we design an ADMM-based distributed algorithm that exploits the problem structure to achieve scalable solutions on standard hardware. Finally, we validate the approach using 1.5\, million real-world trips from Chengdu, demonstrating a 30\% reduction in total cost and improved facility utilization relative to a baseline without assignment.

\section{LLM-Assisted problem modeling and optimization formulation}\label{Optimization}
In this section, we introduce the overall problem setting and optimization framework for EV charging infrastructure planning. The urban area comprises multiple locations that differ in traffic flow intensity, investment cost, and spatial distance, leading to heterogeneous and time-varying charging demand. These factors jointly influence both infrastructure investment and EV assignment decisions. Building on this scenario, we investigate how to allocate charging capacity across locations and assign EVs efficiently, balancing long-term investment and continuous operational objectives. Furthermore, to enhance the formulation process, we propose a human-LLM collaborative framework in which the LLM constructs optimization formulations from textual problem descriptions, while human expertise iteratively guides and refines the process to ensure consistency, interpretability, and validity. The following subsections first analyze the planning problem in detail, then describe the LLM-assisted modeling procedure, and finally present the resulting mathematical formulation.
\begin{figure}[t]
  \centering
   \includegraphics[width=1\linewidth]{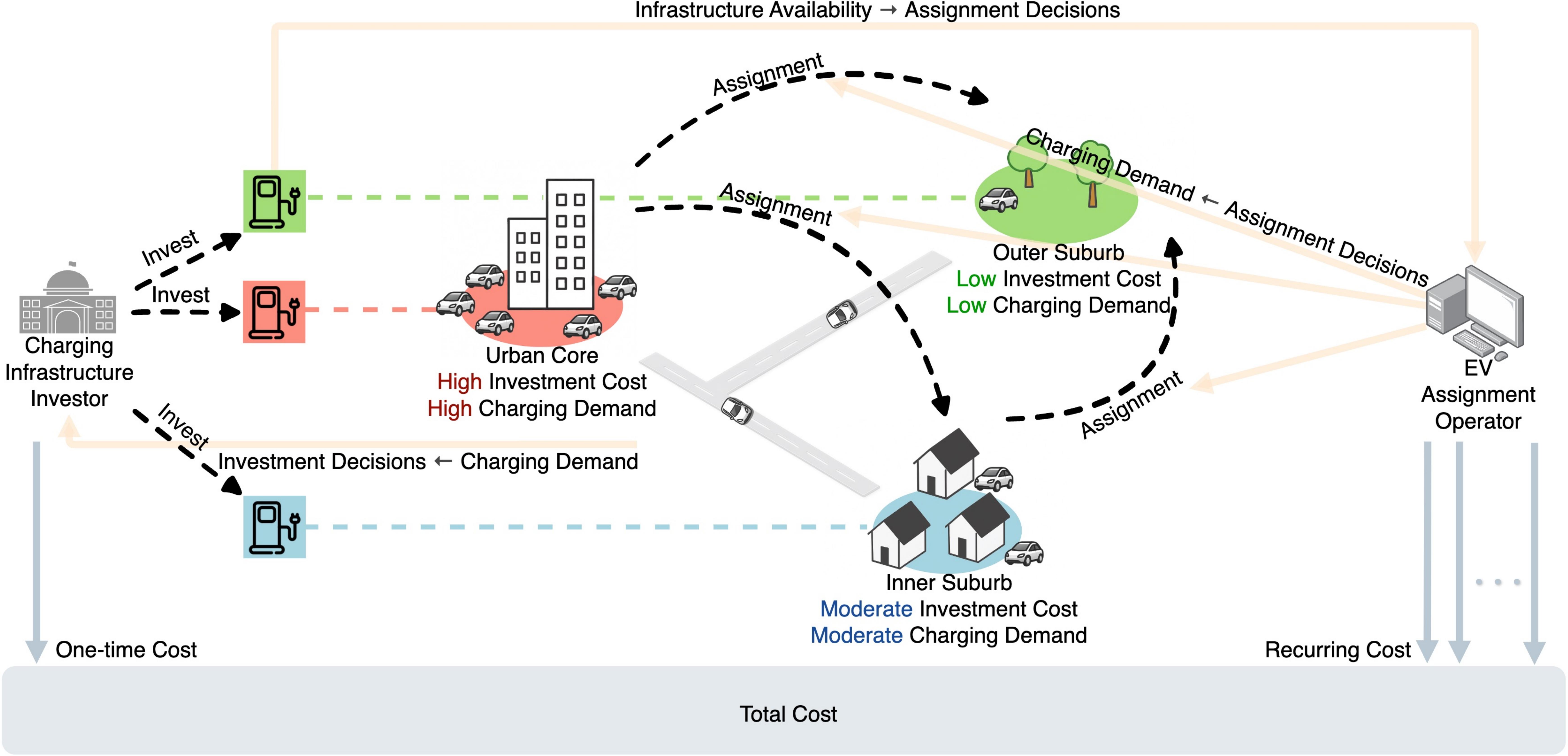}
   \caption{An illustrative example of an EV charging infrastructure planning scenario, showing varying traffic flows, charging demand, and planning costs across urban locations, and a coupled decision framework linking investment and assignment.}
   \label{fig:charging_scenario}
\end{figure}

\subsection{Planning problem analysis}
The EV charging infrastructure planning problem comprises two tightly interconnected components, charging-capacity investment and operational EV assignment, whose mutual dependence induces a spatial-temporal coupling that shapes planning outcomes. As illustrated in Fig.~\ref{fig:charging_scenario}, the urban system contains multiple locations that differ in traffic-flow intensity, land-related investment cost, and pairwise distance. These spatial and temporal variations produce charging demand that varies across locations and over time. Each location’s investment cost is location-specific cost (e.g., land-use price), while operational assignment costs primarily arise from distance-dependent transfers of EVs between charging locations. Consequently, investment and assignment must be treated jointly rather than in isolation.

The decision process involves two roles. The investor selects capacity levels at each location subject to cost-effectiveness and budget constraints; the operator assigns EVs to charging locations in response to real-time demand and available capacity, aiming to minimize the associated operational cost. The two decisions are mutually coupled, as captured in Fig.~\ref{fig:charging_scenario} by the light orange arrows: once the charging capacity at a location is decided, it influences the upper bound in the capacity-satisfaction constraint, limiting the number of EVs that can be feasibly charged in that location (as encoded in constraint~\eqref{capacity_satisfaction}). Conversely, the cumulative assignment patterns across time slots indicate whether the installed capacities are adequate or insufficient. High assignment costs due to long travel distances or frequent reallocation pressures may indicate the need for additional investment at certain sites. This closed-loop interaction between infrastructure availability, assignment behavior, and charging demand underlies the overall planning efficiency and cost.

The planning problem is further complicated by the spatial-temporal coupling of charging demand. Distinct demand patterns arise across different location types (e.g., residential, office, and recreational) and time periods (e.g., morning, noon, and evening), as well as between weekdays and weekends. For example, residential areas typically experience evening peaks, office areas peak around weekday, and recreational areas reach their highest demand during weekends. Consequently, planning based on a single representative period may bias investment toward certain location types while overlooking significant demand peaks elsewhere, resulting in infrastructure underutilization during off-peak hours and potential shortages during local peak periods.

These dynamics call for planning strategies that move beyond static snapshots and instead consider demand patterns holistically. Locations with short but intense demand surges may not warrant large-scale investment, as infrastructure would remain idle for much of the time. In contrast, areas with consistently high demand across multiple time slots, though requiring higher initial investment, can achieve better long-term utilization and cost-effectiveness. Attempts to avoid such investment by reallocating EVs to lower-cost areas may incur higher cumulative assignment costs, offsetting any savings. Furthermore, incorporating spatial-temporal coupling enables planners to stagger demand across both time and space, reducing peak pressure on the grid and improving overall system efficiency.

From the perspective of temporal dimension, the spatial-temporal coupling inherently links long-term investment planning with short-term operational decisions. Investment choices are infrequent and span several years or decades, reflecting strategic commitments that are costly to revise once implemented. By contrast, operational EV assignment decisions occur in short and reduplicative cycles, whose patterns are estimated from historical data, are expected to recur. It is this recurrence that binds the horizons: once the infrastructure is fixed, essentially the same assignment cycles unfold repeatedly under the same structural constraints, and the cumulative performance of these cycles determines the realized adequacy and value of the investment. This nested temporal structure highlights how long-term infrastructure design and recurrent operational behavior jointly shape system performance over time.

Ultimately, the planning system strives to achieve an optimal balance between initial investment expenses in infrastructure and recurrent operational costs associated with EV assignments. Explicitly incorporating the spatial-temporal coupling and mutual interdependencies between these decision-making horizons ensures more effective infrastructure deployment, optimized resource allocation, and promotes sustainable EV infrastructure development. Modeling such a complex, spatial-temporally coupled system manually can be laborious and error-prone, prompting the need for a more efficient modeling paradigm, as explored in the following subsection.

\subsection{LLM-assisted modeling procedure overview}
LLMs are transformer-based neural networks trained on large-scale text corpora to learn statistical patterns in language, capturing both syntactic structures and semantic relationships. This linguistic competence enables them to translate structured natural-language specifications into formal mathematical representations. Given semantically organized descriptions of decision variables, objective functions, and constraints, an LLM can automatically produce symbolic formulations in algebraic or modeling-language-friendly formats (e.g., linear inequalities, summations, index-based expressions). In this way, LLMs facilitate the efficient construction of optimization models directly from domain-level problem descriptions, substantially reducing the reliance on manual formulation in algebraic modeling environments.
\begin{figure}[t]
    \centering
    \includegraphics[width=0.8\linewidth]{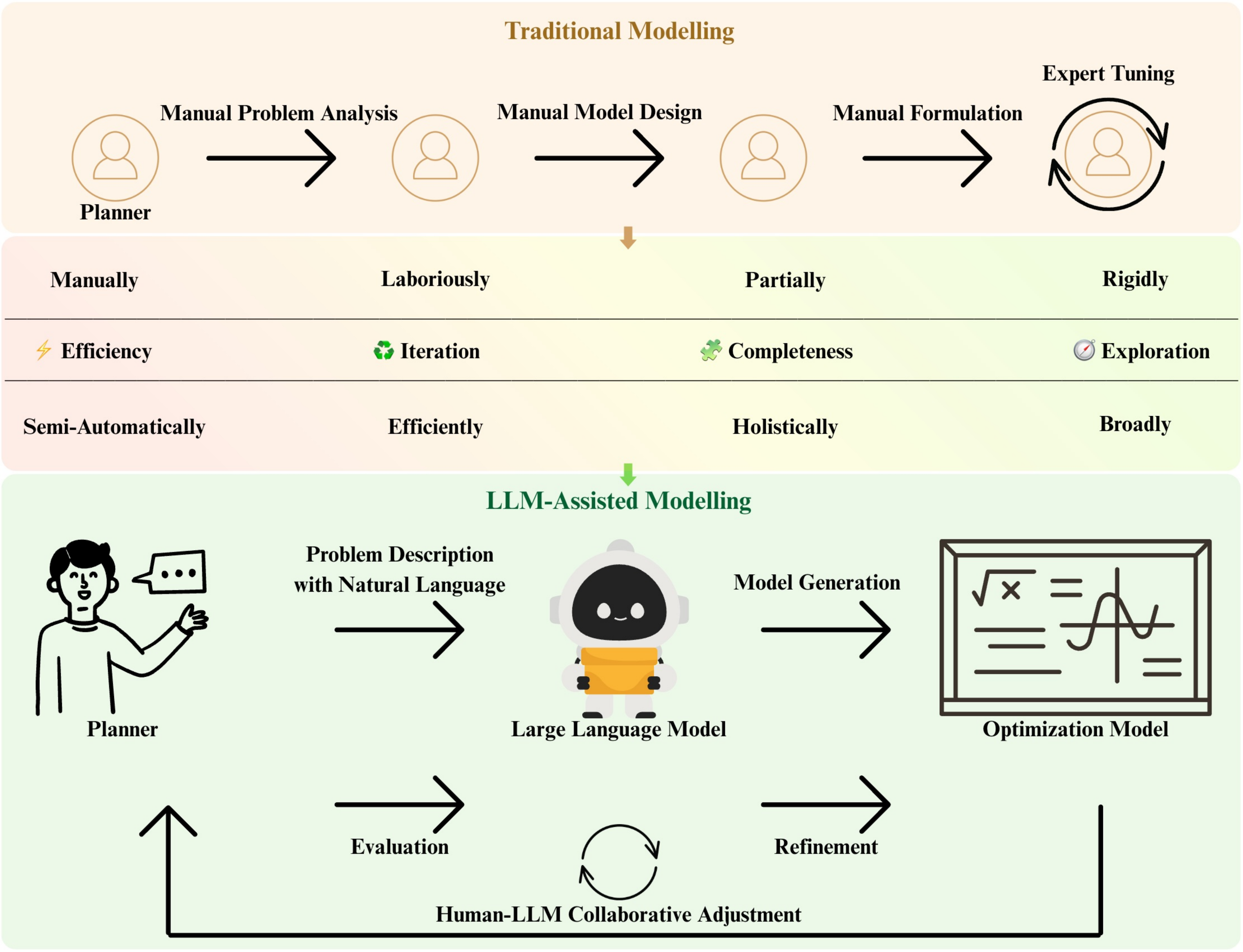}
    \caption{Contrasting traditional and LLM-assisted approaches to the modeling process.}
    \label{fig:llm}
\end{figure}

Building on this capability, we propose an LLM-assisted modeling workflow that redefines traditional model formulation by embedding natural-language interaction into the process (see Fig.~\ref{fig:llm}). In conventional workflows, planners manually analyze problems, design model structures, and construct mathematical formulations step by step, often relying on expert tuning for refinement. This process is typically labor-intensive, fragmented, and rigid, which limits its efficiency and exploratory potential.

In contrast, the proposed workflow integrates human expertise with LLM-based reasoning in an iterative and collaborative loop. The planner begins by describing the planning scenario in structured natural language, outlining decision variables, objectives, and qualitative constraints. The LLM interprets this input and generates an initial mathematical formulation as a draft model. The human expert then reviews and refines the output, identifying missing constraints, unrealistic assumptions, or ambiguous notation. Through successive exchanges, where human judgment guides machine generation, the formulation iteratively converges toward both mathematical soundness and domain relevance.

This human-LLM co-modeling paradigm fundamentally transforms the modeling landscape. Compared with traditional methods that proceed manually, laboriously, partially, and rigidly, the LLM-assisted approach operates semi-automatically, efficiently, holistically, and broadly. The transformation manifests in faster formulation, enhanced iterative flexibility, greater structural completeness, and richer exploratory potential. Collectively, these advantages enable planners to generate high-quality optimization models with significantly reduced manual effort, marking a shift from problem-solving through explicit derivation to collaborative knowledge synthesis between humans and intelligent systems.

While the LLM facilitates model development and helps structure complex optimization problems, its role is assistive rather than autonomous. The generated outputs are not guaranteed to be correct or complete and must be reviewed and refined by human experts to ensure mathematical validity and contextual relevance. In this collaborative framework, the LLM acts as a modeling assistant, supporting rapid prototyping and iteration, while deferring critical modeling decisions to domain experts.

The reliability of LLM-assisted modeling is grounded in a human-in-the-loop validation process embedded throughout the interaction. Rather than deferring correctness checks to a post hoc stage, model components are incrementally verified and refined during generation. This stepwise validation enables the detection of structural inconsistencies, ensures alignment with domain constraints, and maintains semantic coherence. As such, correctness in this workflow emerges not from the LLM's autonomous capacity, but from a collaborative modeling dynamic in which expert judgment continuously shapes and safeguards the formulation. In summary, validation in our LLM-assisted modeling workflow relies on three key mechanisms that ensure the correctness and contextual relevance of the generated models while maintaining expert oversight throughout the process:

\begin{itemize}
    \item \textbf{Prompt specification}: Clear and well-specified prompts are used to accurately convey modeling intent, guiding the LLM to generate relevant and interpretable mathematical components.
    
    \item \textbf{Iterative human-in-the-loop refinement}: Human experts interact with the LLM during model development, validating and modifying intermediate outputs to ensure semantic correctness and alignment with modeling objectives.
    
    \item \textbf{Final expert validation}: The completed model is reviewed by experts to verify mathematical soundness, completeness, and consistency with problem-specific requirements.
\end{itemize}

\subsection{LLM-assisted optimization model formulation}
Building upon the LLM-assisted modeling approach described previously, we now introduce the explicit mathematical formulation of the joint investment and operational EV assignment optimization model. The detailed structure and constraints presented here directly reflect the spatial-temporal characteristics and interdependencies discussed earlier, and were iteratively refined through interactions with the LLM. In this work, we apply the GPT-4o model provided by OpenAI \cite{openai_chatgpt}.

The three scenarios in Fig.~\ref{fig:model-dialogue} demonstrate how the LLM-assisted workflow transforms the traditional modeling process through interactive reasoning and rapid iteration. In scenario (a), the workflow begins with a structured natural-language description of the planning problem. The user specifies the planning objective, decision scope, and spatial-temporal context of EV charging infrastructure. From this input, the LLM automatically formulates a mixed-integer linear program, defining decision variables, objective function, and constraints. This initial draft provides a mathematically coherent foundation that captures the user’s intent, turning a conceptual problem statement into a formal optimization model.

Scenario (b) illustrates how the LLM facilitates rapid ideation and exploratory reasoning. After the expert notes that traffic flow cannot directly represent charging demand, the LLM reformulates the relationship by introducing a stochastic conversion factor \(\alpha_i^t \in [0,1]\) and enumerates candidate distributions to represent its uncertainty. This exchange broadens the expert’s reasoning space, quickly producing multiple mathematically valid alternatives that accelerate conceptual exploration.

Scenario (c) highlights the workflow’s strength in rapid refinement. When the expert points out that the initial formulation implicitly assumes instantaneous transfers, the LLM amends the temporal constraints by introducing a travel-delay parameter \(\tau_{ji}\), ensuring that EVs dispatched from \(j\) contribute to demand at \(i\) only after arrival. Through such iterative dialogue, the model is corrected and strengthened without restarting from scratch-combining domain insight with computational adaptability to achieve efficient, realistic model development.

Together, these examples demonstrate the complementary strengths of human and LLM contributions. The LLM is able to quickly produce complete draft formulations and, once provided with targeted feedback, to efficiently revise the model with mathematically consistent adjustments. Meanwhile, human experts inject contextual insights and highlight domain-specific nuances. This interaction shows that the workflow is not only capable of avoiding subtle modeling oversights, but also of accelerating the overall modeling process, reducing the manual burden while improving realism and accuracy. This reduces the time from problem description to first working model from hours to minutes.

Below, we present the final optimization model derived through the LLM-assisted formulation process, which are finally refined and validated by the authors as optimization and energy experts. It is important to note that the formulation presented below is not the product of a single automatic generation. Instead, it reflects a sequence of human-LLM interactions, where experts progressively guided the LLM to refine its initial drafts into a rigorous and contextually appropriate model. This collaborative process highlights the main advantage of LLM-assisted modeling: the ability to combine automated generation with expert knowledge for rapid yet reliable model construction.

\textit{Charging infrastructure investment model.} From the investor's standpoint, the objective is to determine the optimal charging capacity investment to maximize overall benefits. Simultaneously, the invested charging capacity must meet the charging demand at each location and each moment after assignment. The investment cost is determined by the charging capacity demand and the cost per unit capacity at each location. The corresponding decision is represented as $\boldsymbol{c}_i = \{c_i, \forall i \in \mathcal{I}\}$, where $c_i$ denotes the invested charging capacity at location $i$. This decision indicates the required charging capacity to be installed at location $i$. If $c_i = 0$ for any $i$, it implies that no charging station is needed at location $i$. The cost per unit capacity consists of a base investment cost $\pi_0$ and a location-dependent investment cost $\pi_i$ for location $i$, in which $\pi_0$ includes both the base cost covering installation and materials, and $\pi_i$ denotes the location-specific cost, such as land use cost for accommodating the invested charging capacity. The objective function representing the total cost for the investment problem is formulated as
$$
\sum\limits_{i \in \mathcal{I}} c_i \times (\pi_0 + \pi_i),
$$
where the goal is to minimize the total investment cost, which is defined as the cumulative base costs $\pi_0$ and location-dependent cost $\pi_i$ incurred from investments in charging capacities $c_i$ across all designated locations $i$. The investment needs to satisfy the following constraints:
\begin{align}
    & \sum\limits_{i \in \mathcal{I}} c_i \times (\pi_0 + \pi_i) \leq B ,\label{total_budget}\\
    & 0 \leq c_i \leq C^{max}_i, \forall i. \label{max_capacity}
\end{align}
Constraint \eqref{total_budget} is associated with a maximum investment cost subject to a total budget of $B$, with a lower bound of 0 for investment cost. Constraint \eqref{max_capacity} specifies the upper and lower bounds for the charging capacity at each location, where the maximum capacity is determined by location-dependent factors, such as electricity grid, and the minimum capacity should be non-negative. A charging capacity cannot be negative, and a value of 0 signifies no requirement for charging station investment.

\textit{EV charging assignment model.} The operator's role is to determine where the EVs can be charged in a cost-effective and convenient way based on the available capacity at each site. The operator may assign EVs to alternative charging locations, incurring associated costs, primarily dictated by the travel distances when assigning EVs between locations. From the operator's viewpoint, the goal is to determine the best EV assignment strategy to minimize these costs.

The assignment costs are calculated based on the number of EVs assigned and the distance between two locations for the assignments. For any pair of locations $i$ and $j$, we define $z^t_{ij}$ as the number of EVs assigned from location $i$ to location $j$ in time slot $t$. The cost for each assignment is determined by the product of the number of assignments $z^t_{ij}$ and the assignment cost per EV $L_{ij}$ between the two locations correspondingly, where $i, j \in \mathcal{I}$. For computational efficiency, the assignment variable $z^t_{ij}$ is relaxed to be continuous. Since $z^t_{ij}$ represents the number of vehicles assigned in aggregated cohorts, treating it as continuous is common practice in traffic assignment and flow-based formulations. Integer values can be recovered by rounding. The objective function representing the total cost for the assignment problem is formulated as
$$
\sum\limits_{t \in \mathcal{T}}
\sum\limits_{i \in \mathcal{I}}\sum\limits_{j \in \mathcal{I}} z^{t}_{ij} \times L_{ij},
$$
where the goal is to minimize the total cost resulting from EV assignment costs across all locations and time slots. EV charging and assignments need to satisfy the following constraints
\begin{align}
        & 0 \leq \sum\limits_{j \in \mathcal{I}} z^{t}_{ij} \leq \alpha^t_i \cdot F^{t}_i, \forall{i,t},\label{flow_conservation}\\
        & 0 \leq \beta \cdot (\alpha^t_i \cdot F^t_i - \sum\limits_{j \in \mathcal{I}} z^{t}_{ij} +\sum\limits_{j \in \mathcal{I}} z^{t-\tau_{ji}}_{ji}) \leq c_i, \forall{i,t},\label{capacity_satisfaction}\\
        & z^{t}_{ij} \ge 0, \forall{i, j}, \label{positive}\\
        & z^t_{ij} = 0, \forall i = j, \label{local_assignment}\\
        & z^t_{ij} = 0, \forall i,j, \text{if }L_{ij} \ge R. \label{range}
\end{align}
Constraint (\ref{flow_conservation}) specifies that the number of EVs assigned from location $i$ to $j$ must be no more than the original traffic flow at location $i$ requiring charging, with $\alpha^t_k$ denoting the estimated coefficient between charging demand and traffic flow at location $k$ during time $t$. Constraint (\ref{capacity_satisfaction}) ensures that the charging capacity at each location is sufficient (i.e., greater than or equal to the charging demand) in order to meet the total charging demand across all time slots, where the charging demand is calculated by incorporating EV assignments. The coupling between long-term investment and short-term assignment is explicitly captured in such a constraint. For example, the charging capacity variable $c_i$ appears in the capacity constraint~(\ref{capacity_satisfaction}), ensuring that EV assignment flows at each time slot cannot exceed the installed infrastructure capacity. At the same time, the assignment variables $z^t_{ij}$ aggregate across time slots and locations to determine the effective utilization of charging infrastructure, thereby influencing the optimal allocation of $c_i$. Thus, the constraints jointly encode how investment limits operational feasibility, while operational flows retroactively reveal the adequacy and efficiency of investment decisions.

The coefficient $\beta$ represents the relationship between the number of EVs requiring charging and the charging capacity demand, describing the connection between investment in charging infrastructure and charging demand. To accurately capture the arrival dynamics of EVs assigned across locations, we introduce a delay-aware constraint. Specifically, vehicles assigned from location $j$ to $i$ require $\tau_{ji}$ time slots to arrive. Thus, the available charging demand at i in time slot t accounts for inflows originating at $t - \tau_{ji}$.  Given that variables $z^{t}_{ij}$ and $z^{t}_{ji}$ respectively represent the flows in opposite directions, the sign of these variables is not required to be considered. Hence, Constraint \eqref{positive} mandates that these assignments must be non-negative, ensuring that the flow values are appropriately restricted to reflect actual capacities and prevent computational anomalies associated with negative allocations. Constraint \eqref{local_assignment} prevents the model from assigning vehicles to their original location to avoid causing non-convexity issues, as the distance between a location and itself is 0. Lastly, Constraint \eqref{range} stipulates that the distance of assignments must be within an acceptable range, denoted as $R$.

\textit{Joint investment and assignment model.} Considering the interdependence between investment and assignment, the investment and operation are integrated by incorporating a one-time investment cost and the expected recurring assignment costs over the entire investment horizon. Specifically, the operation costs are multiplied by the expected recurrences $E_t$ for each corresponding time slot $t$, where $E_t$ is a set of numbers. The term ``expected recurrence'' implies that we anticipate a similar scenario in a time slot $t$ occurring $E_t$ times throughout the investment horizon. The goal is to minimize the global investment costs for each location $i$, as well as the expected operational costs for all locations $i$ and all time slots $t$ over $E_t$ expected recurrences. Therefore, the joint investment and assignment optimization problem is formulated as
\begin{equation}
\begin{aligned}
 \min_{c_i, z^{t}_{ij}} & \sum\limits_{i \in \mathcal{I}} c_i \times (\pi_0 + \pi_i) \\
& + \sum\limits_{t \in \mathcal{T}}
\sum\limits_{i \in \mathcal{I}} \sum\limits_{j \in \mathcal{I}}  E_t \times z^{t}_{ij} \times L_{ij},\\
\text{s.t.}& ~\eqref{total_budget}-\eqref{range}.
\end{aligned}
\end{equation}
In summary, the final formulation is a linear programming (LP) model, with both objective and constraints linear. The total is minimized subject to all constraints across both the investment and operational horizons. The decision variables are $c_i$ and $z^t_{ij}$, corresponding to the optimal charging capacities and EV assignment decisions, respectively. The overarching goal is to achieve a globally optimized solution by minimizing the overall investment and assignment costs. This bidirectional linkage is explicitly embedded in the joint formulation, where investment decisions set the feasible capacity region and assignment decisions determine the realized utilization and costs. As such, the constraints serve as the mechanism through which long-term investment limits short-term operational feasibility, while short-term operational flows retroactively inform the adequacy and efficiency of long-term investment decisions.

\section{Distributed solution method via the ADMM}\label{ADMM}
As previously noted, the temporal-spatial variation in charging demand must be thoroughly considered to guarantee precise and dependable outcomes. Nevertheless, as more data are incorporated into the optimization process, the computational burden grows substantially. Moreover, when considering larger geographic regions, data privacy emerges as a potential concern for governmental bodies. A distributed algorithm can mitigate the extent of information sharing required, including sensitive data pertaining to investment budgets, capacities of power systems, and traffic flows. Consequently, we design a distributed algorithm leveraging the ADMM algorithm.

Given the complex dimensionality of the decision variables for EV assignments, particularly as more locations or time slots are incorporated, central optimization becomes less practical. Postulating a scenario involving $n$ locations results in $n^2$ location pairs. If $m$ time slots are taken into consideration, there will be $n^2 \cdot m$ EV assignment decisions to be made. Therefore, this poses a substantial challenge for optimal decision computation. To overcome this computational challenge, we can decompose the computation of EV assignment decisions. In the proposed distributed algorithm, each location independently optimizes decisions regarding its own charging capacity and the assignment of EVs to other locations, treating them as parallel sub-problems. Subsequently, the algorithm only adjusts decisions related to charging capacities in a master problem and then relays the outcomes back to each sub-problem to enable result adjustments.

Specifically, based on ADMM, we introduce auxiliary variables $\tilde{c_1}, \cdots, \tilde{c_n}$ for \( c_1, \ldots, c_n \), and the Lagrangian multipliers (dual variables) $\lambda_i$ for $c_i - \tilde{c_i} = 0$, and we define $z^t_i = [z^t_{i,1}, z^t_{i,2},..., z^t_{i,j}]$, where $j \in \mathcal{I}$ indicating the assignment decisions made by location $i$ to location $j$. Then, we derive the augmented Lagrangian function as
\begin{equation}
    \begin{aligned}
        \mathcal{L} = \sum_{i \in \mathcal{I}} \sum_{t \in \mathcal{T}} & f_i(z^{t}_{i}, c_i) + \lambda_i(\tilde{c_i} - c_i) +  \frac{\rho}{2} (\tilde{c_i} - c_i)^2,\\
    \end{aligned}
\end{equation}
where $\rho > 0 $ is the penalty parameter, and the objective function for each location $i$ is 
\begin{equation}
    \begin{aligned}
        f_i(z^{t}_{i}, c_i) = c_i \cdot (\pi_0 + \pi_i) + E \cdot \sum_{j \in \mathcal{I}}z^t_{i,j} \cdot L_{ij}.\\
    \end{aligned}
\end{equation}

Since the formulated optimization problem is a LP, it is convex, and the ADMM algorithm applies with its theoretical guarantees. Based on the augmented Lagrangian function, we divide the optimization problem into sub-problems of locations and a master problem of center. In sub-problems, each location optimizes its primal variables $c_i$ and independent variables $z^t_i$ to minimize their investment and assignment costs. Then, locations submit their variables $c_i$ and  $z^t_i$ to the center while the center updates the auxiliary variables $\tilde{c_i}$ and dual variable $\lambda_i$. The algorithm iterates to update the auxiliary variables and dual variables until it converges.

In each location's sub-problem, the cost is minimized in parallel by taking the auxiliary variable $\tilde{c_i}$ and dual variable $\lambda_i$ as given. The sub-problem is formulated as
\begin{equation}\label{eq:AL}
    \begin{aligned}
        \min_{c_i, z^t_i} & \sum_{t \in \mathcal{T}} f_i(z^{t}_{i}, c_i) - \lambda_i \cdot c_i +  \frac{\rho}{2} (\tilde{c_i} - c_i)^2,\\
        \text{s.t.}& ~\eqref{total_budget}-\eqref{range}.
    \end{aligned}
\end{equation}

Note that the primal investment-assignment problem is an LP (linear objective, affine constraints) and hence convex. For fixed multipliers, the augmented Lagrangian in Eq.~\eqref{eq:AL} is convex in the primal variables (linear terms + convex quadratic penalty + indicator of an affine set). With the chosen splitting, each ADMM update solves a convex subproblem.

Recalling the constraints outlined in the centralized optimization problem, the distinction emerges: each location imposes constraints solely on its own decisions. Notably, Constraint \eqref{capacity_satisfaction} specifies that the decision regarding charging capacity must fulfill the resultant charging demand, which is based on the original traffic flow deducted by the EV assignments to other locations and augmented by assignments received from other locations. It is imperative to note that a location cannot determine the assignments it receives from other locales. Such EV assignments are provided as parameters by the overarching master problem, denoted as $z'_i$. The $z'_i$ is made by transformation according to the EV assignment decisions $z_i$ made from the previous iteration, and it does not require any calculation.

Besides updating the auxiliary variables $\tilde{c_i}$ and the dual variables $\lambda_i$, the center manages the capacity satisfaction constraint for all locations. For the center's master problem, it obtains primal variables $c_i$ and also the variables $z^t_i$ from all location $i \in \mathcal{I}$, and then calculates the auxiliary variables $\tilde{c_i}$ through the master problem as

\begin{equation}
\begin{aligned}
\min_{\tilde{c_i}} & \sum\limits_{i \in \mathcal{I}}[\lambda_i(\tilde{c_i} - c_i) + \frac{\rho}{2} (\tilde{c_i} - c_i)^2],\\
& \text{s.t.}~\eqref{capacity_satisfaction}.
\end{aligned}
\end{equation}

\begin{figure}[t]
  \centering
   \includegraphics[width=0.6\linewidth]{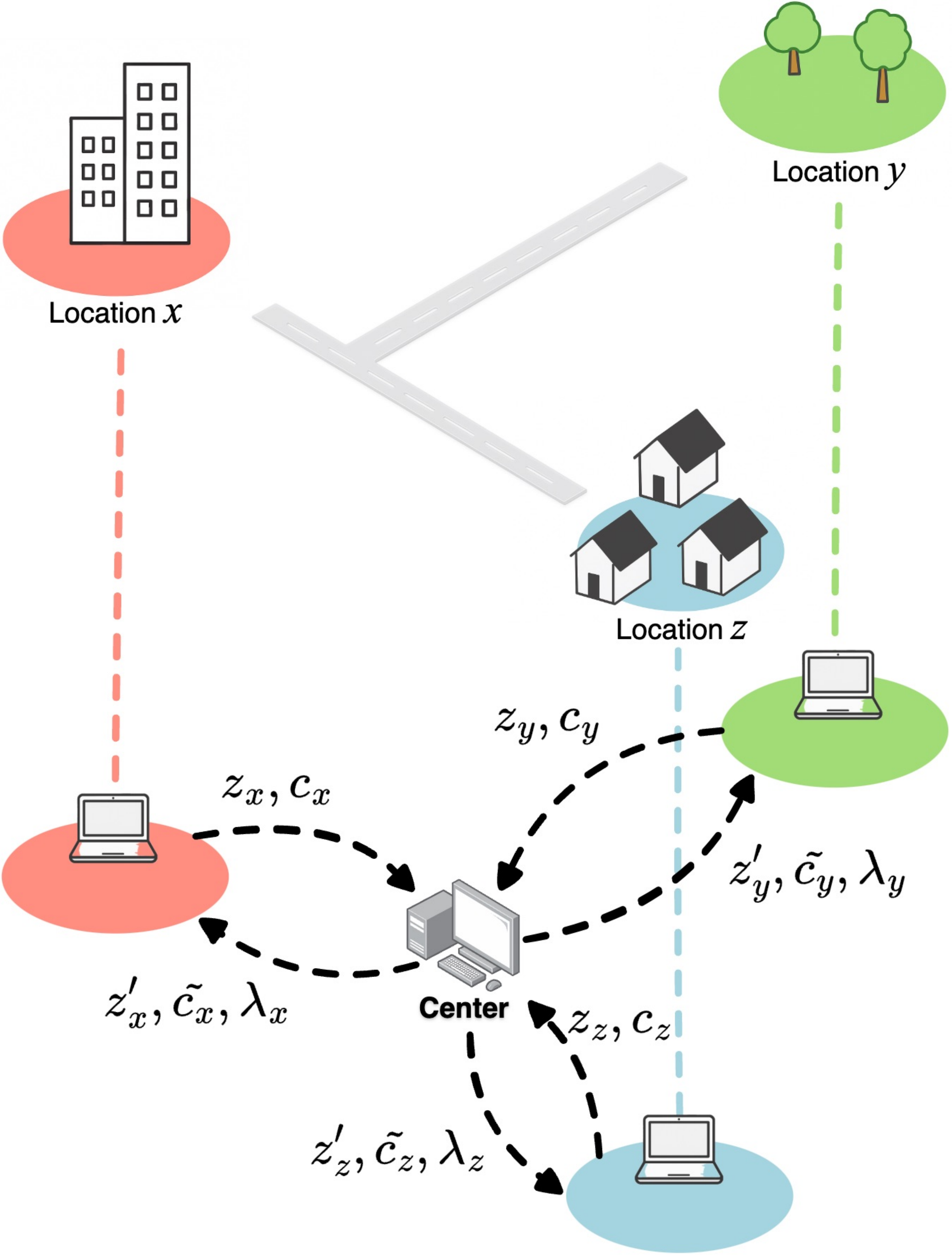}
   \caption{An illustrative example of the computing and communication process for distributed ADMM.}
   \label{fig:admm_scenario}
\end{figure}

The master problem needs to satisfy Constraint \eqref{capacity_satisfaction}, which ensures the satisfaction of charging capacity for every location. Overall, the decision variables and auxiliary variables are exchanged between each location and a computational center, as illustrated in Fig. \ref{fig:admm_scenario}.

For each iteration $k$, a solution set $(c^k_1, ..., c^k_n)$ and the corresponding Lagrangian multipliers $(\lambda^k_1, ..., \lambda^k_n)$ are determined. Subsequently, the ADMM advances to the subsequent iteration, denoted as $k + 1$, as
\begin{align}
    &  \lambda_i[k+1] = \lambda_i[k] + \rho\cdot(\tilde{c}[k] - c[k])\label{iteration_admm}.
\end{align}

The algorithm concludes its operations once a predetermined condition, such as a maximum number of iterations or a convergence within a threshold, is met. The quality of the solution at each iteration can be assessed through the primal error and dual error as
\begin{align}
    &  Q_{primal} = \sum\limits_{i \in \mathcal{I}}|\tilde{c}_i[k] - c_i[k]|\label{primal_error},\\
    &  Q_{dual} = \sum\limits_{i \in \mathcal{I}}|\lambda_i[k] - \lambda_i[k-1]| \label{dual_error}.
\end{align}
When both errors attain a value of zero, it indicates that the ADMM has converged to an optimal solution. Overall, the optimization solution approach can be implemented as Algorithm \ref{admm_algorithm}.

\begin{algorithm}[t]
\DontPrintSemicolon
\caption{Distributed ADMM algorithm for EV charging infrastructure planning} \label{admm_algorithm}
\textbf{Search parameters:} $\rho$ , $k_{max}$, $Q_{threshold}$ \;
 $\lambda_i, C, Z, \tilde{C}, Z', Cost, Q_{primal}, Q_{dual}, k \leftarrow$ Initialize()\;
\While{$Q_{primal} > Q_{threshold} \mid Q_{dual} > Q_{threshold}$}{
    \For{$i \in \mathcal{I}$}{
        \textbf{Sub-problem:} \;
        $c_i$, $z^t_i$, $Cost_i$ $\leftarrow$ Optimize($\tilde{C}[i]$, $Z'$, $\lambda_i[k]$)\;
        $C[i]$,  $Z[i]$, $Cost[i]$ $\leftarrow$ Update($z^t_i$, $c_i$, $Cost_i$ )\;
        }
    $Z'$$\leftarrow$ Transform($Z$)\;
     \textbf{Master problem:} \;
     $\tilde{C}$$\leftarrow$ Optimize($C$, $Z$, $\lambda$)\;
     \textbf{Lagrange multiplier update:} \;
    $\lambda_i[k+1] \leftarrow \lambda_i[k] + \rho\cdot(\tilde{C}[k] - C[k])$ \;
    \textbf{Calculate errors in $k$-th iteration:} \;
    $Q_{primal}, Q_{dual} \leftarrow \text{Update}$()\;

    $k$ $\leftarrow$ $k + 1$\;
    \If{$k$ $\ge$ $k_{max}$}{
    \textbf{ break}
    }
}
\end{algorithm}

\section{Case study using real-world data}\label{Case study}
To demonstrate the efficacy of our proposed model and evaluate the potential benefits it can provide for EV charging infrastructure planning with EV assignment, we conduct an in-depth case study on a joint investment and operation system using real-world data from Chengdu, China.

\subsection{Data processing and optimization model parameters}

The data employed in this study are sourced from the ``Didi Chuxing GAIA Initiative" \cite{didi}, an anonymized dataset provided by Didi Chuxing Technology Co., Ltd-a mobile transportation platform. This dataset contains 1.5 million realistic travel records from Chengdu, China, gathered between November 21 and 27, 2016. November in Chengdu represents a typical working month without major holidays \cite{china_briefing_holidays_2016} or extreme weather conditions \cite{timeanddate_chengdu_2016}, which makes it relatively stable for baseline analysis. The available data capture marked variability within a weekly operational cycle, revealing distinct weekday–weekend contrasts in both temporal intensity and spatial distribution of charging demand, such as concentrated peaks around business districts on weekdays and broader dispersion toward residential and recreational areas during weekends. We partition Chengdu into $201$ distinct sub-districts for optimization purposes, each representing a unique location. The dataset incorporates travel time and destination data, expressed in longitude and latitude coordinates, which are subsequently used to calculate the initial traffic flow $F^t_i$ for each location $i$ and time slot $t$. We adopt one week as the operational horizon and a ten-year investment horizon, implying 520 expected recurrences; accordingly, the recurrence factor $E_t$ is set to 520.

\subsubsection{Charging demand}
Traffic flow data in OD-matrix form are processed to derive time-varying traffic volumes across different districts, with origins and destinations identified from their geographic coordinates. The data are then organized into 15-minute time slots over seven days of a week, resulting in 672 distinct time slots. Adopting a 15-minute time scale offers greater precision in operations and captures more detailed fluctuations in charging demand over time. The data are then aggregated into 201 distinct locations. A zero value is imputed for cases where no traffic flow is observed for a specific time and location. This process yields a dataset comprising 135,072 ($672 \times 201$) rows, enabling the determination of traffic flow distribution across each location and time slot. This distribution serves as the foundation for simulating Chengdu's overall traffic flow and is utilized as the initial vehicular flow $F^t_i$ for the optimization. 

The coefficients $\alpha^t_i$ in Constraints \eqref{flow_conservation} and \eqref{capacity_satisfaction} represent the relationship between the traffic flow at time slot $t$ in location $i$, and the number of traffic flows requiring charging. To represent the proportion of traffic flow requiring charging (denoted as $\alpha$ in \eqref{capacity_satisfaction}), we implement it using a Beta distribution which is defined on the interval $[0, 1]$ and well-suited for proportion-type variables. By adjusting the shape parameters, the Beta distribution can flexibly represent various charging demand patterns and incorporate prior knowledge or spatial-temporal variability. For the case study, we set the Beta distribution parameters to $\text{Beta}(10, 90)$, which assumes an average charging demand of 10\% with moderate variability across different locations and time periods, ensuring smooth and stable simulation outcomes. The coefficient $\beta$ in \eqref{capacity_satisfaction} represents the relationship between the number of traffic flows requiring charging and the demand for charging capacity. In this context, we refer to \cite{tesla}, the foremost EV brand in today's market, and adopt a fast charging power of 250 kW as the value of $\beta$. This choice is based on the expectation that fast charging will become pervasive in the future, specifically over a ten-year time horizon.

\subsubsection{Distances}
The distances between two locations are calculated by the average distance from all the records. We believe that the dispatch distance between two locations cannot be represented by the straight distance between two locations because the trajectory of a trip is rarely straight between two points. Moreover, a sub-district cannot be simply assumed as a point. However, real-world traffic data can provide information on which points in the region are where traffic congregates or which points are more likely to be the start and end points. We obtain the actual assignment distance by summarizing all the records of each pair of locations, taking the average value and generating the distances for optimization. The acceptable range $R$ of assignment is set to 3 km, since we considered drivers are generally willing to tolerate only short detours to access charging facilities in urban contexts \cite{WANG2024186}. According to \cite{ev_cost}, the assignment cost is estimated at $0.2$ Chinese Yuan per unit distance. To more accurately represent the traffic road network topology within the context of model constraints, we define the assignment cost per EV $L_{ij}$ in the optimization as
\begin{equation*}
    L_{ij} = \left \{
    \begin{aligned}
        & \text{$\infty$},~ \text{if}~\text{distance(}i,j\text{)} \ge R, \\
        & 0,~ \text{if}~ i = j, \\  
        & \text{distance(}i,j\text{)} \cdot 0.2,~\text{otherwise},
    \end{aligned}
   \right.
\end{equation*}
where distance($i$, $j$) is the average travel distance between location $i$ and location $j$.

\subsubsection{Investment cost and budget}
The investment cost comprises a base cost and a location-dependent cost per unit of charging capacity. The base cost per unit capacity $\pi_0$ is estimated at $500$ Chinese Yuan for rapid charging, as calculated by the price of the charging pile from \cite{tesla}. The location-dependent cost per unit capacity $\pi_i$ is approximated based on the distance between the location and the center of Chengdu, Chunxi Street, which has the highest property prices in the city \cite{ceic}. Consequently, we assume that Chunxi Street, as Chengdu's city center, possesses the highest location-dependent cost per unit capacity. Utilizing Chunxi Street as a reference point, we decrease the location-dependent costs for other locations following an exponential distribution to generate the respective costs.

According to the Chengdu EV charging and battery swapping infrastructure specialized planning by Chengdu Municipal Bureau of Economy and Information Technology \cite{chengdu_ev_infra_2023} and the estimated investment cost for a charging station \cite{CHEN2023125747}, this case study estimates a maximum budget $B$ of 20 billion Chinese Yuan over a ten-year investment horizon. Note that this is a synthetic study, and the budget is adjustable.

\subsection{Optimization results analysis}
This section illustrates the convergence of the proposed ADMM algorithm and the performance compared to the centralized method. Also, the insights on optimal planning resulted from the proposed model considering EV assignment compared to a basic flow-capture model. The optimization solving process is performed on a computer equipped with an Apple M1 Pro CPU, 16 GB random access memory (RAM), and the macOS system. 

\subsubsection{Performance of proposed ADMM algorithm}

\begin{figure}[t]
  \centering
   \includegraphics[width=0.9\linewidth]{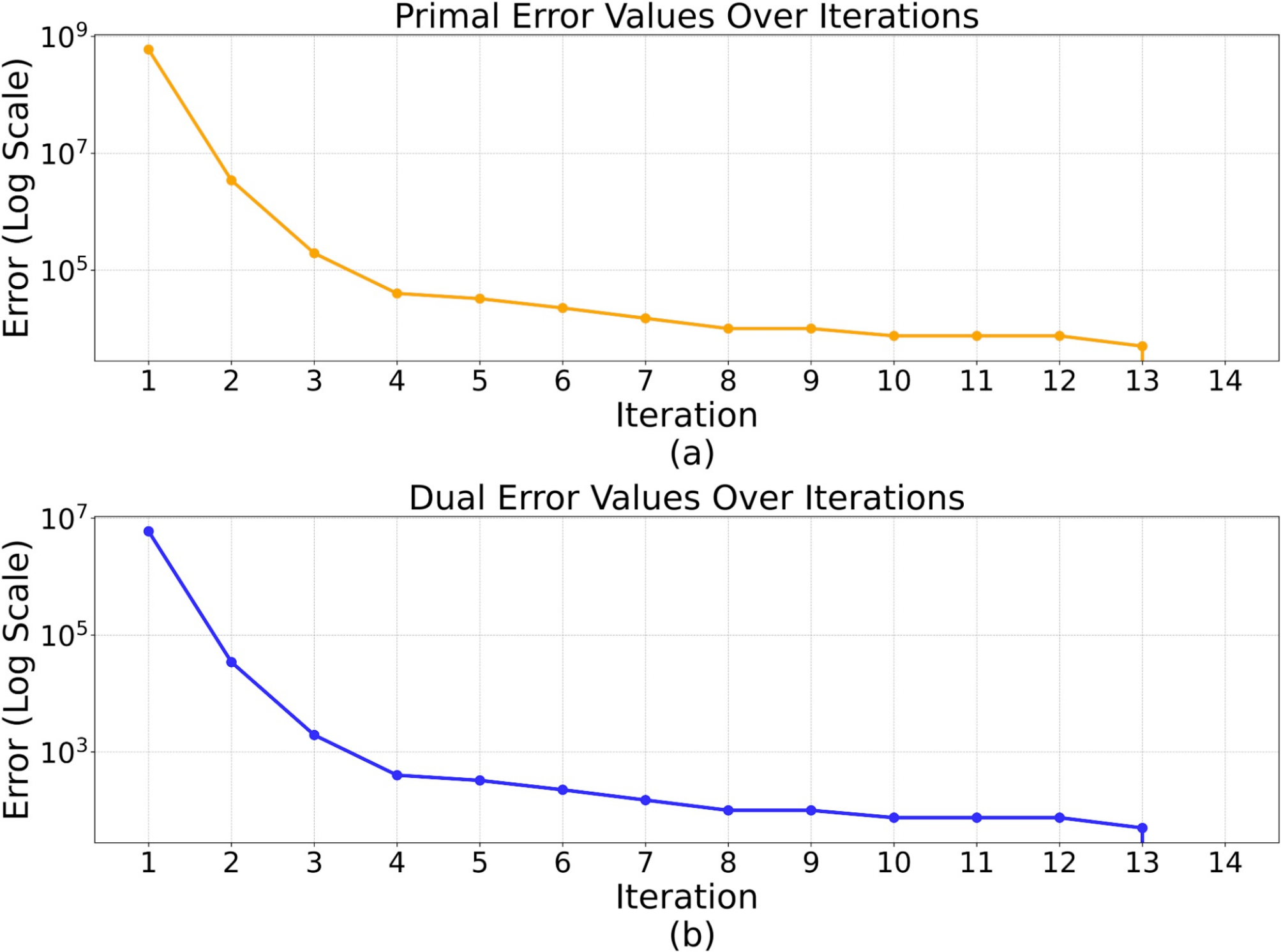}
   \caption{Primal and dual errors over iterations for the ADMM approach.}
   \label{fig:error}
\end{figure}

We validate the proposed ADMM algorithm in the case study. As described in Algorithm \ref{admm_algorithm}, we iteratively set every location to solve for their own optimal decision as a single sub-problem and use the master problem to converge to a globally optimal solution. Fig.~\ref{fig:error} presents the primal and dual errors over iterations, and the errors are in a logarithmic scale. The result shows that the algorithm is able to converge efficiently to an optimal solution in 14 iterations for our problem. The main parameter settings used in our experiments as follows: the penalty parameter $\rho$ was set to 0.1, the convergence threshold $Q_{threshold}$ was set to $10^{-4}$. These values follow common practice in \cite{Boyd_Parikh_Chu_Peleato_Eckstein_2010} and were found to provide stable and efficient convergence in our case study. For completeness, we note that the decision variables were assigned practical bounds to ensure feasibility and numerical stability during optimization. Specifically, the upper bound of charging capacities $C_i$ was set to $10^7$ and the upper bound of assignment flows $Z_{ij}$ was set to $10^4$, while the lower bounds were set to zero. These bounds are sufficiently large so as not to restrict the feasible solution space, but they prevent the algorithm from exploring unrealistic or numerically unstable regions.

For comparison, we also centrally solve the optimization problem as a baseline. We employ the \cite{gurobi} optimization solver to solve the optimization problem centrally at once to identify the optimal global solution for the infrastructure investment decisions for every location and their EV assignments for every time slot. The centralized approach demands 50 minutes of computational time. However, this computational process is associated with an excessively high memory consumption of 76.31 GB of RAM. Owing to the memory swapping management mechanism inherent to the macOS, the procedure is able to reach completion despite the substantial memory demands. However, this level of computational burden likely represents the upper threshold of what a standard computer can manage, particularly given the data span only one week. 

The distributed optimization employing the ADMM method starkly contrasts with the centralized approach. The distributed method markedly improves efficiency, with each sub-problem and the corresponding master problem being solved in under one second. We also employ the Gurobi optimization solver in each sub-problem and master problem for calculating the optimal decision and auxiliary variables. Each sub-problem can be solved independently in parallel in an iteration. Presuming the operation within a truly distributed computing environment and discounting the latency attributable to data communication, it is feasible to complete each iteration in two seconds. Consequently, the optimization problem can be resolved in an estimated 30 seconds over the course of 14 iterations. This represents nearly a 100-fold reduction in computation time compared to the approximately 50 minutes required by the centralized solver. Moreover, the memory footprint for each sub-problem is markedly reduced, necessitating less than 500 MB of RAM. This modest memory consumption and computational burden facilitate the practicality of the distributed method for deployment on most commercial computers.

Comparing outcomes, the distributed approach incurs a total cost of 15,744 million Chinese Yuan, which is marginally higher than the 15,674 million Chinese Yuan achieved via the centralized method, representing an acceptable error margin of approximately 0.45\% in total cost.
\subsubsection{Insights on optimal results}
\begin{figure*}[!t]
  \centering
  \begin{subfigure}{0.45\linewidth}
    \includegraphics[width=.9\textwidth]{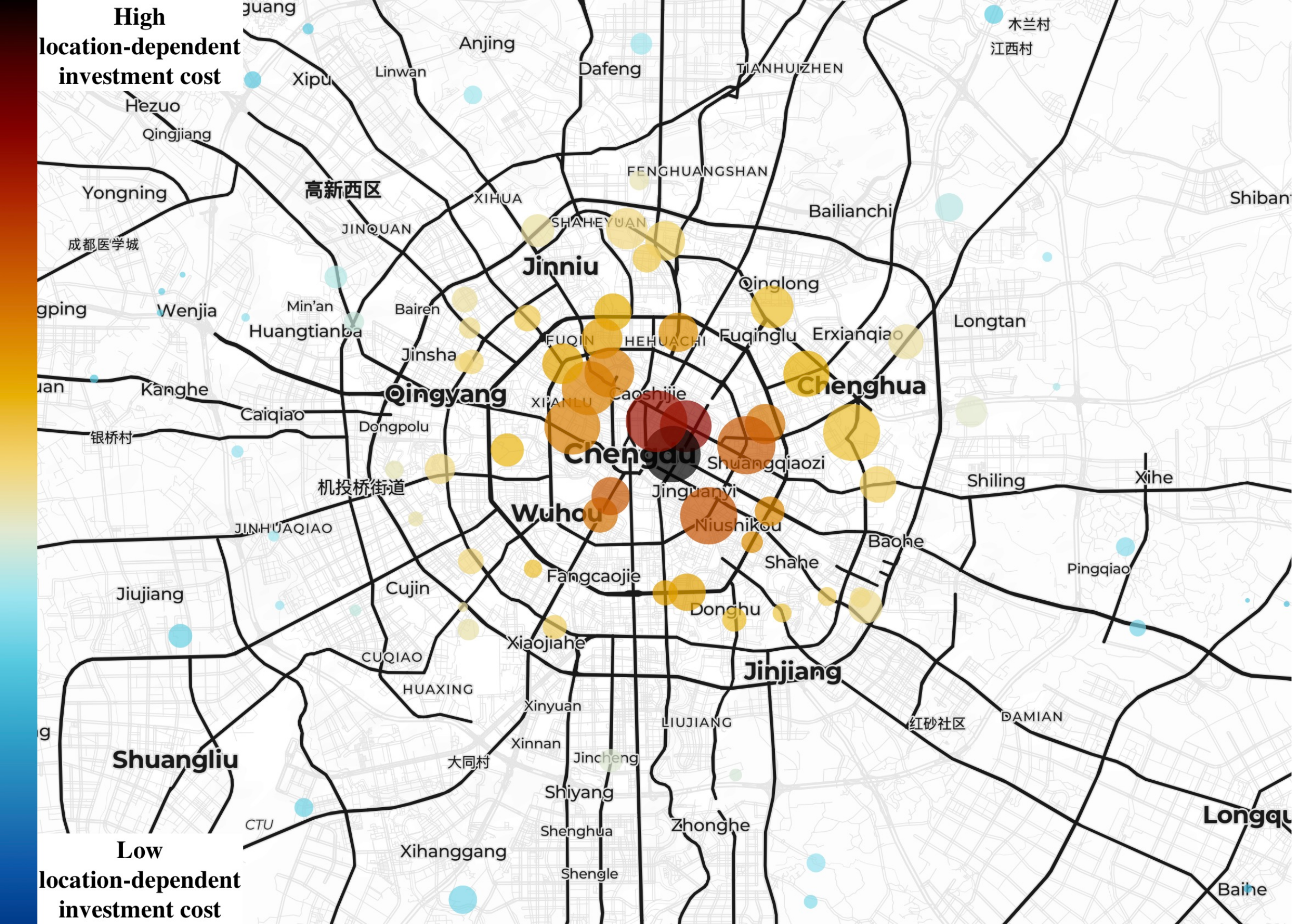}
    \caption{Base model.}
    \label{fig:base}
  \end{subfigure}
  \hfill
  \begin{subfigure}{0.45\linewidth}
    \includegraphics[width=.9\textwidth]{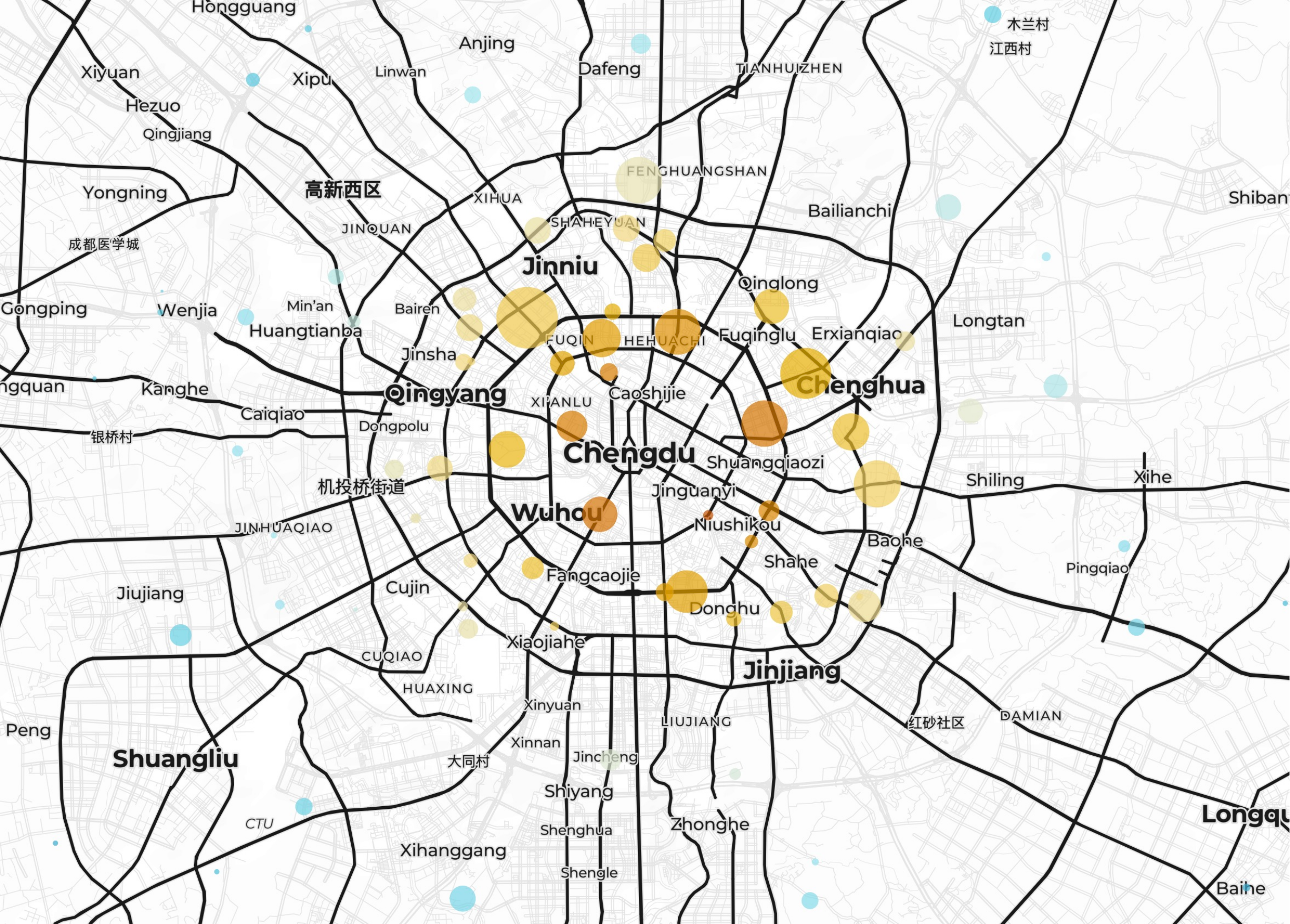}
    \caption{Proposed model.}
    \label{fig:model}
  \end{subfigure}
  \caption{Optimal investment decisions in maps showing the distribution of the investment decisions. The bubbles correspond to the locations where the area of the bubbles represents the amounts of charging capacities. The area of bubbles of the two models has been separately scaled according to the proportion of total investment decisions. The color represents the location-dependent investment cost for different locations. }
  \label{fig:result}
\end{figure*}

Fig.~\ref{fig:result} presents the visualizations of the investment decisions of capacity for the base model and our proposed model in each location. Fig. \ref{fig:base} illustrates the base model, which optimizes investment costs independently, without considering assignment. In this model, charging infrastructure planning employs a basic flow-catching model, minimizing capacity while fulfilling the charging demands for all locations and time slots. The base model shows that charging capacity should be more concentrated in the city center area with high charging demand, without incorporating assignment. In contrast, Fig. \ref{fig:model} portrays the optimal solution derived from the proposed model, which incorporates assignments. This model accounts for the interrelated effects between investment and assignment and arrives at mutually advantageous decisions. The comparison between the two figures highlights a substantial alleviation of the excessive load in the central area when considering EV assignment. Most charging investments are therefore located in the densely developed inner urban zones surrounding the core, where charging demand remains high but land costs are still manageable. The proposed model effectively reallocates part of the investment from the congested center to these adjacent areas, achieving a balanced configuration that enhances cost efficiency without causing excessive dispersion of charging infrastructure into low-demand suburban regions. Upon closer observation, it is apparent that investment decisions in specific suburban regions are substantially lower compared to those derived from the baseline model, even resulting in no investment in some of these areas, such as the locations around Wenjia (left-hand side of the map). As previously mentioned, when a specific location experiences transient traffic flow, assigning this ephemeral traffic flow to another region for charging purposes can yield enhanced overall benefits.

\begin{table*}[h!]
\centering
\caption{Comparison of costs under different assignment ranges $R$. 
The base model corresponds to $R=0$ (no assignment flexibility). The Cost Reduction column reports the percentage reduction in total cost relative to the immediately smaller $R$ (i.e., compared with the previous case). All cost values are expressed in million Chinese yuan (CNY).}
\label{tab:R_sensitivity}
\begin{tabular}{ccccc}
\toprule
$R$ (km) & 
Investment Cost & 
Assignment Cost & 
Total Cost & 
Cost Reduction (\%) \\
\midrule
 0 & 22,464 & 0 & 22,464 & -- \\
\midrule
 1 & 18,834 & 15 & 18,849 & 16.09 \\
 \textbf{3} & \textbf{15,516} & \textbf{228} & \textbf{15,744} & \textbf{16.47} \\
 5 & 14,201 & 425 & 14,626 & 7.10 \\
 7 & 13,760 & 455 & 14,215 & 2.81 \\
 9 & 13,663 & 467 & 14,130 & 0.60 \\
 11 & 13,568 & 493 & 14,061 & 0.49 \\
\bottomrule
\end{tabular}
\label{tab:numerical}
\end{table*}

From a numerical perspective, as summarized in Table~\ref{tab:numerical}, the base model yields an optimal objective of approximately 22,464 million CNY, whereas the proposed joint investment–assignment model with a 3 km assignment range achieves 15,744 million CNY, corresponding to a cost reduction of about 30\%. Over the ten-year period, the accumulated assignment cost amounts to only 228 million CNY, confirming that modest operational adjustments can lead to substantial overall economic benefits. Furthermore, Table~\ref{tab:numerical} illustrates how varying the assignment range $R$ affects system performance. Increasing $R$ allows EVs to access charging stations over longer distances, thereby improving infrastructure utilization and reducing the required investment. As a result, the total system cost decreases with larger $R$, though the reduction is non-linear. The marginal benefit diminishes once most reassignable demand has been absorbed-beyond a certain threshold, further expansion of $R$ yields limited additional savings, since capacity bottlenecks are largely relieved while assignment costs from extended travel distances continue to accumulate.

\begin{figure*}[!th]
\centering
    \begin{subfigure}{0.45\linewidth}
        \includegraphics[width=.9\linewidth]{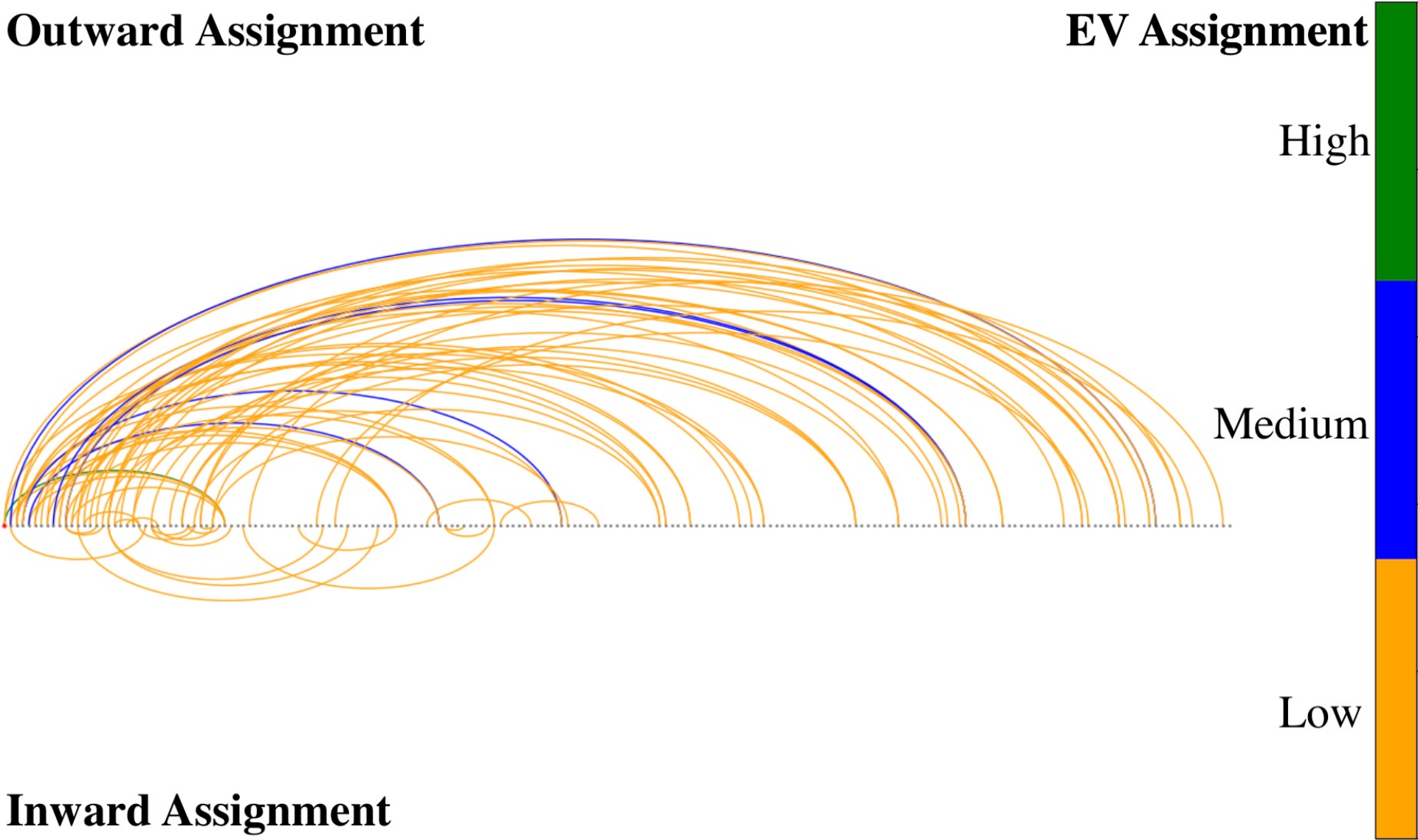}
    \caption{Arc plot for Monday 3:00 - 4:00 PM.}
    \label{fig:z_a}
    \end{subfigure}
\hfil
    \begin{subfigure}{0.45\linewidth}
        \includegraphics[width=.9\linewidth]{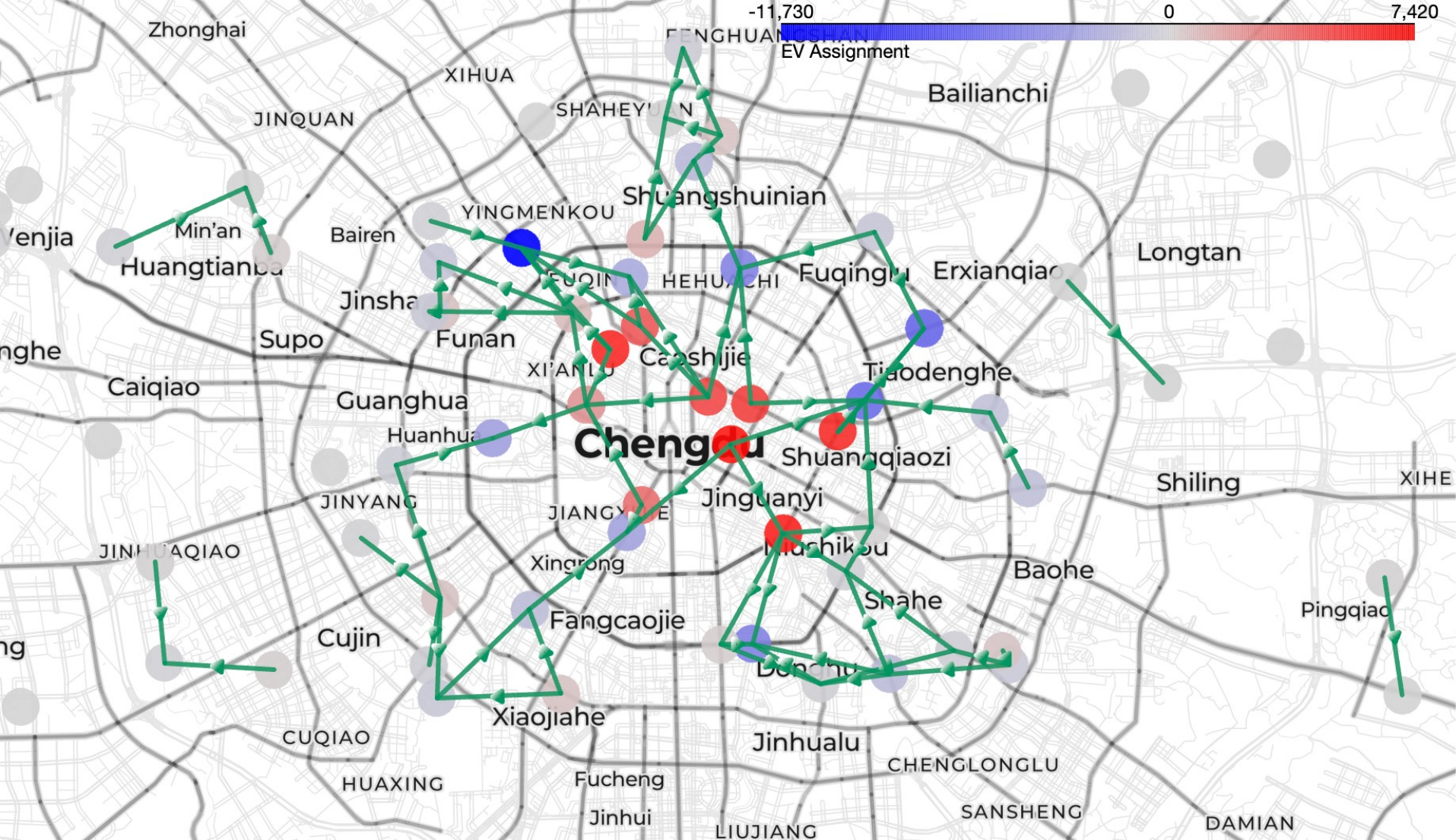}
    \caption{Map for Monday 3:00 - 4:00 PM.}
     \label{fig:z_b}
    \end{subfigure}

    \begin{subfigure}{0.45\linewidth}
        \includegraphics[width=.9\linewidth]{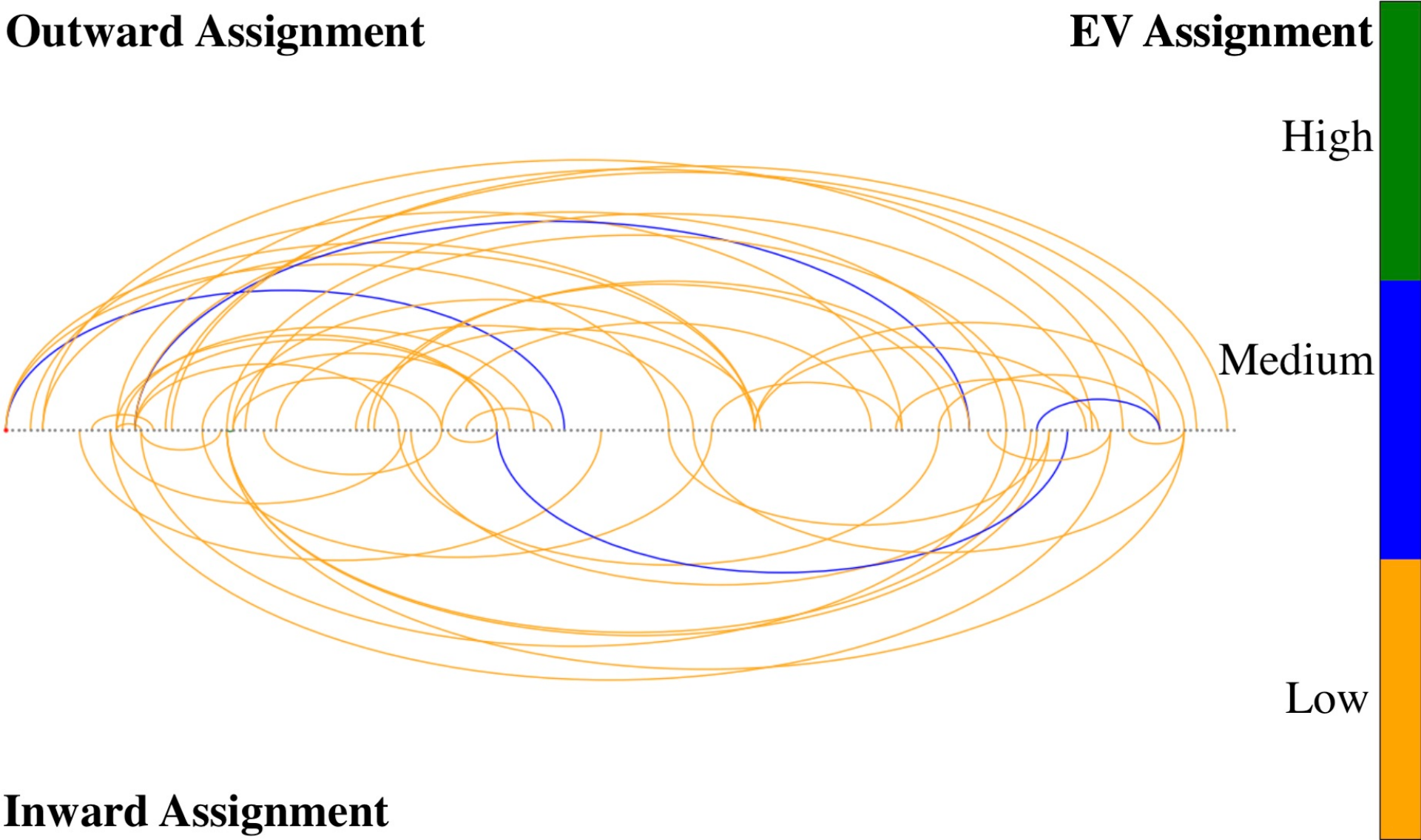}
    \caption{Arc plot for Monday 10:00 - 11:00 PM.}
     \label{fig:z_c}
    \end{subfigure}
\hfil
    \begin{subfigure}{0.45\linewidth}
        \includegraphics[width=.9\linewidth]{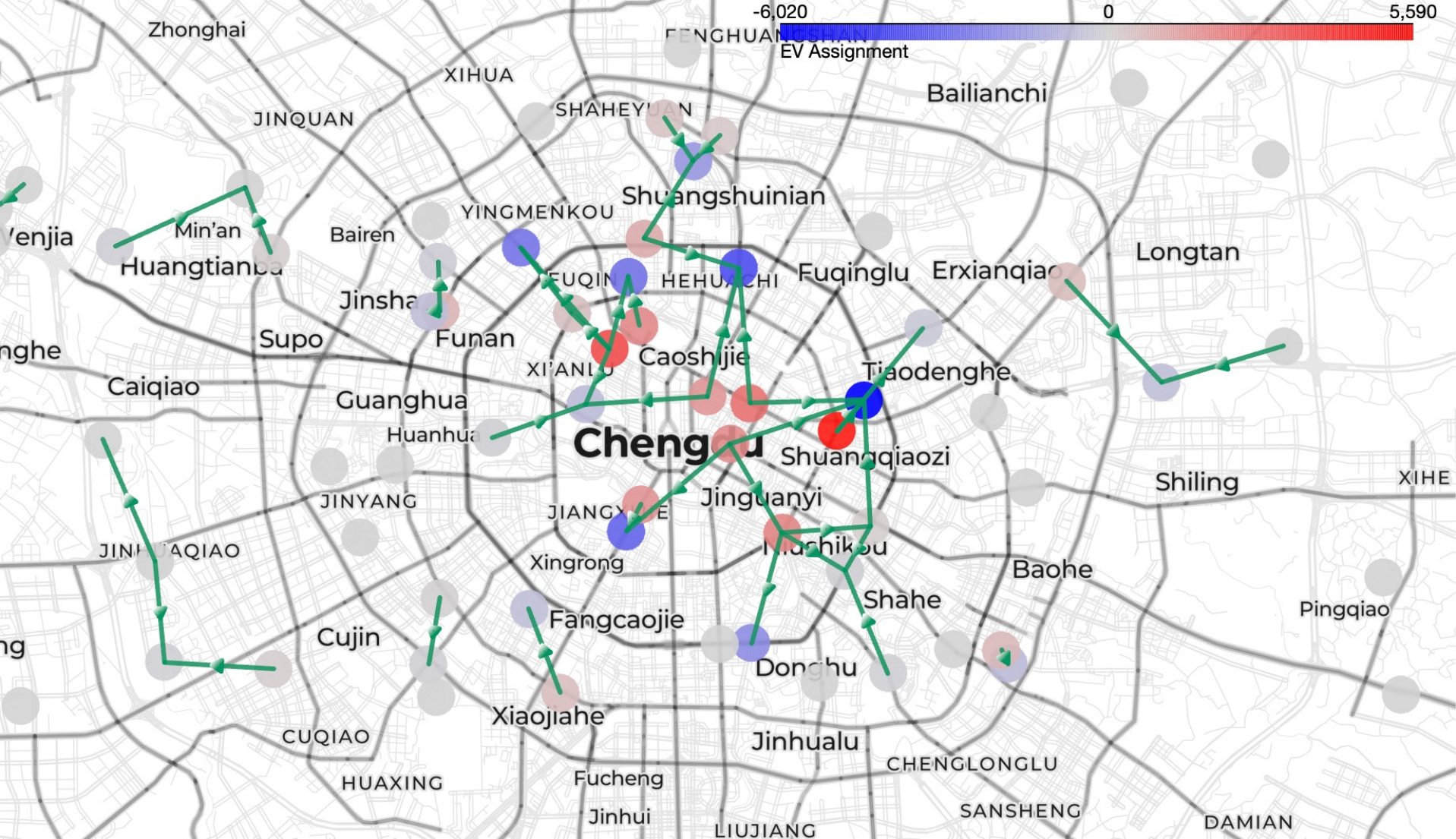}
    \caption{Map for Monday 10:00 - 11:00 PM.}
     \label{fig:z_d}
    \end{subfigure}

    \begin{subfigure}{0.45\linewidth}
        \includegraphics[width=.9\linewidth]{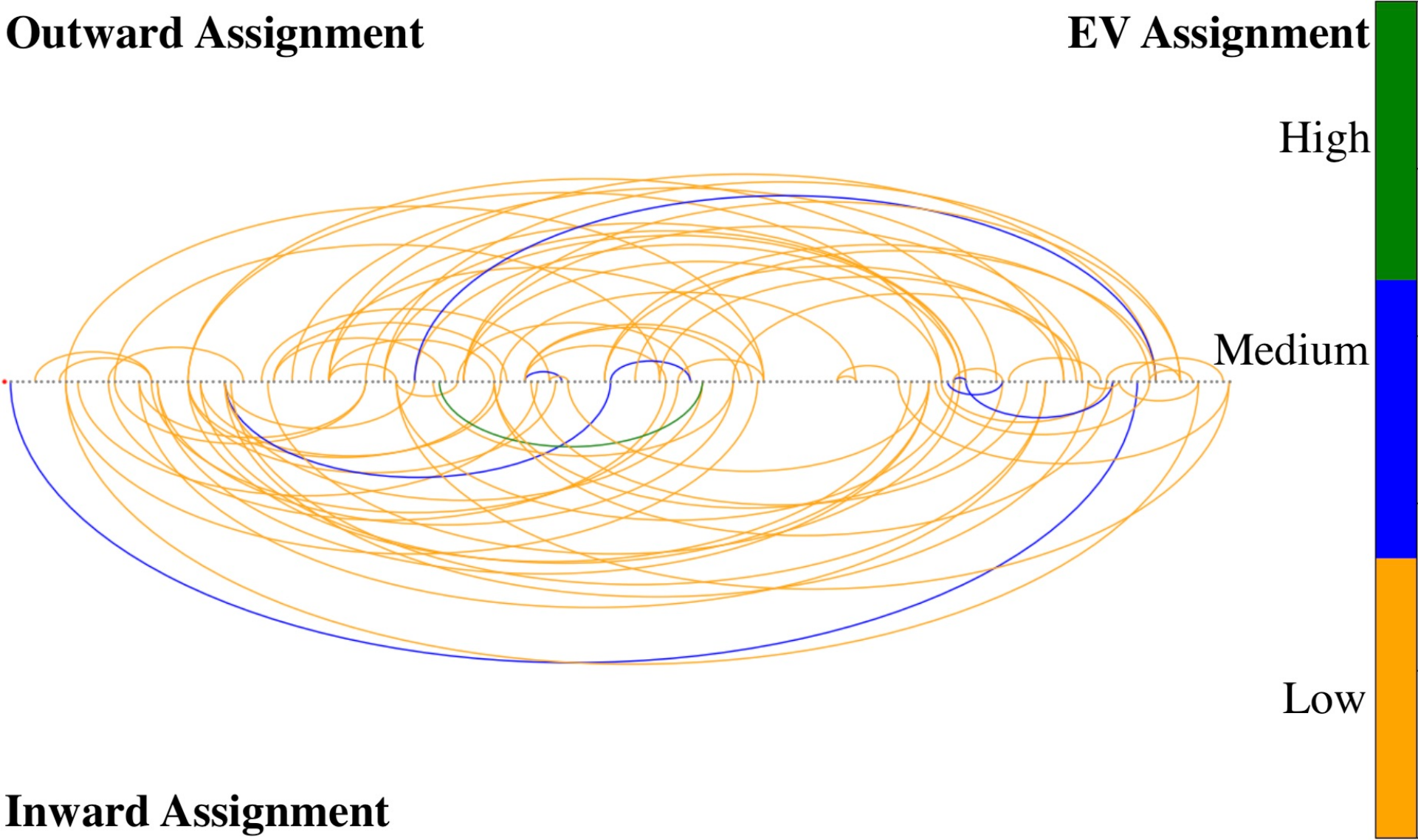}
    \caption{Arc plot for Saturday 3:00 - 4:00 PM.}
     \label{fig:z_e}
    \end{subfigure}
\hfil
    \begin{subfigure}{0.45\linewidth}
        \includegraphics[width=.9\linewidth]{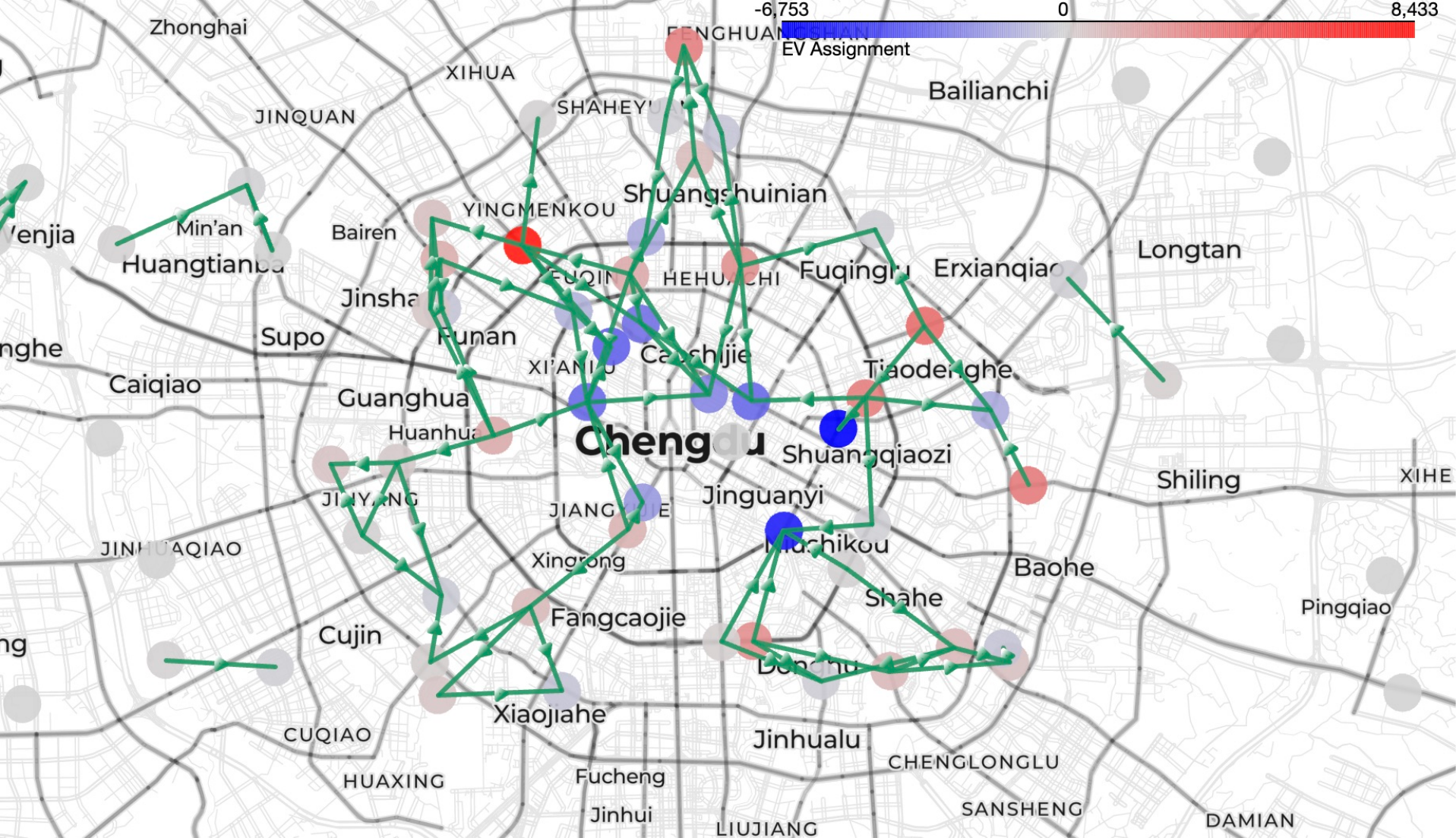}
    \caption{Map for Saturday 3:00 - 4:00 PM.}
     \label{fig:z_f}
    \end{subfigure}

\caption{EV assignment decisions snapshot for different periods. For (a), (c), and (e), EV assignment decisions are represented in arcs between different locations sorted by the investment cost (also refer to the sorting from the city center to the suburb), and the numbers of EV assigned are divided into three levels. The EV assignments from the center to the suburbs are shown at the top, while the EV assignments from the suburbs to the city center are shown at the bottom. For (b), (d), and (f), EV assignment decisions are represented in maps. Each bubble corresponds to a location, with the color of the bubble indicating the number of EV assignments. A deeper color signifies a higher number of assignments. As the term ``number of EV assignments" refers to the number of EVs assigned to another location, while a negative value indicates EVs received from other locations. The lines display the direction of the EV assignments.}
    \label{fig:EV_assignment}
    \end{figure*}

The proposed model provides EV assignment decisions for all time slots and locations, enabling a comprehensive understanding of the charging infrastructure planning process. To obtain a deeper understanding of the temporal regularity in decision patterns, we analyze snapshots captured at multiple intervals and compare the outcomes produced by our model. Fig. \ref{fig:EV_assignment} depicts the patterns of EV distribution across three distinct time periods, highlighting the temporal variation in assignment strategies. First, the depicted maps in Figs.\ref{fig:z_b}, \ref{fig:z_d}, and \ref{fig:z_f}, affirm that all EV assignments align with the practical constraints of our model, underscoring its applicability to real-world scenarios. 

Specifically, Fig. \ref{fig:z_a} presents an Arc plot visualization of EV assignment distribution between 3:00-4:00 PM on a Monday, revealing a pronounced trend where a majority of EVs are assigned from urban centers to suburban areas. Further analysis, as illustrated in Fig. \ref{fig:z_b}, shows the EV assignments around and within the city center during the same period, indicating a directional flow of EVs outward from the city center. 

In contrast, during the later hours of the same day, between 10:00-11:00 PM (as shown in Fig. \ref{fig:z_c}), there is a discernible shift in the assignment dynamics. This period sees a reduction in EV dispatches from urban to suburban areas and an increase in the reverse flow, highlighting the efficient utilization of charging infrastructure and underscoring the importance of considering the spatial-temporal variation in charging behavior. Furthermore, Fig. \ref{fig:z_d} illustrates a decrease in overall EV assignments during this timeframe, albeit with consistent origins and destinations for EV flows. This contrasts with the patterns observed during 3:00-4:00 PM, where Fig. \ref{fig:z_b} indicates a reduction in intensity at central locations. 

Differences in assignment behaviors are not only evident across different times of the day but also across days of the week. Fig. \ref{fig:z_e} highlights the disparity between weekdays and weekends, showing a notable inclination for EVs to move from suburban areas back to urban centers on Saturdays, the opposite of the Monday trend. This variation, further visualized in Fig. \ref{fig:z_f}, where central locations are marked in blue to signify the reception of EVs, confirms the presence of spatial-temporal coupling affecting EV distribution. 

Intuitively, EVs are expected to be assigned from areas with higher investment costs to those with lower costs to minimize overall expenditure. However, Fig.~\ref{fig:z_c} and \ref{fig:z_e} show that a considerable number of EVs are assigned from the outskirts toward the city center, which may initially appear counterintuitive. This pattern arises because the model fully accounts for temporal coupling, capturing time-varying charging demand across different periods and locations. Consequently, the model assigns part of the suburban charging demand to central areas to achieve globally optimal costs when the charging load in high-cost regions is relatively low. This strategic assignment underscores the importance of integrating temporal and spatial dimensions in EV charging infrastructure planning, as it enables more efficient resource utilization and a clearer understanding of the interaction between investment costs and charging demand.

\begin{figure*}[th]
  \centering
\tcbset{llmchatbox/.style={
  colback=gray!5!white,
  colframe=blue!0!black,
  fonttitle=\bfseries,
  coltitle=black,
  boxrule=0.3pt,
  arc=1mm,
  left=2mm, right=2mm, top=1mm, bottom=1mm
}}

\begin{tcolorbox}[llmchatbox, title=-]
\footnotesize

\textbf{Scenario (a) - initial model formulation from natural language}

\textbf{\textcolor{green!50!black}{User:}} I’m planning EV charging infrastructure across a city. Each location has setup costs (fixed and location-dependent), and EV charging demand varies by time and location. I want to determine how much capacity to install and how to assign EVs in each time slot to minimize combined investment and travel costs, capturing the spatial-temporal coupling. Could you formulate this as a mathematical optimization model, including decision variables, objective function, and constraints?

\vspace{0.6em}
\textbf{\textcolor{blue!60!black}{GPT-4o:}} Certainly! Here’s a mathematical optimization model for planning EV charging infrastructure, minimizing both investment and operational (travel) costs. This is a mixed-integer linear program (MILP).

\textbf{Objective:}
\[
\min_{c_i, z^t_{ij}} \sum_{i \in \mathcal{I}} c_i \cdot (\pi_0 + \pi_i) + \sum_{t \in \mathcal{T}} \sum_{i \in \mathcal{I}} \sum_{j \in \mathcal{I}} E_t \cdot z^t_{ij} \cdot L_{ij}
\]

\textbf{Sets and Indices:}
\begin{itemize}[leftmargin=2em]
  \item[$\bullet$] \( i \in \mathcal{I} \): Charging locations.
  \item[$\bullet$] \( t \in \mathcal{T} \): Time slots.
\end{itemize}

\textbf{Decision Variables:}
\begin{itemize}[leftmargin=2em]
  \item[$\bullet$] \( c_i \in \mathbb{R}_{\ge 0} \): Charging capacity installed at location \( i \).
  \item[$\bullet$] \( z^t_{ij} \in \mathbb{Z}_{\ge 0} \): Number of EVs assigned from location \( i \) to location \( j \) at time \( t \).
\end{itemize}

\textit{...... (Additional constraints and parameters definitions omitted for brevity.)}

\medskip\noindent\color{gray!60}\rule{\linewidth}{0.5pt}\color{black}\medskip

\textbf{Scenario (b) - stochastic representation of charging demand}

\textbf{\textcolor{green!50!black}{User:}} 
Traffic flow cannot be directly equated to charging demand. Only part of the traffic actually needs charging. Since the exact conversion ratio is hard to observe, let’s treat it as a random variable within $[0,1]$. List several possible distributions for this stochastic coefficient.

\vspace{0.6em}
\textbf{\textcolor{blue!60!black}{GPT-4o:}} Under this setting, the effective charging demand can be expressed as:
\[
D_i^t = \alpha_i^t \cdot F_i^t
\]
where $\alpha_i^t \in [0,1]$ denotes the stochastic conversion factor capturing the proportion of traffic requiring charging. Several candidate distributions can be considered for $\alpha_i^t$:  
\begin{itemize}
    \item \textbf{Truncated normal distribution}: restricted normal distribution to a finite interval, ensuring all realizations lie within $[0,1]$.  
    \item \textbf{Logit-normal distribution}: applying the logistic transformation to a normal variable, yielding outcomes strictly to $[0,1]$.
    \item \textbf{Beta distribution}: a flexible two-parameter distribution defined on $[0,1]$, capable of representing various shapes from uniform to highly skewed.
    \item \textbf{Uniform distribution }: the simplest baseline assumption on $[0,1]$, representing complete uncertainty about the conversion ratio.

\end{itemize}
If empirical data exist, an empirical or mixture form can be considered.

\medskip\noindent\color{gray!60}\rule{\linewidth}{0.5pt}\color{black}\medskip

\textbf{Scenario (c) - resolving temporal inconsistency}

\textbf{\textcolor{green!50!black}{User:}} When assigning EVs from location \( j \) to location \( i \) at time \( t \), it takes time for the vehicles to travel from \( j \) to \( i \). However, it appears that the demand at location \( i \) in time slot \( t \) is directly increased without accounting for this travel time.

\vspace{0.6em}
\textbf{\textcolor{blue!60!black}{GPT-4o:}} Great observation. The formulation assumes instantaneous EV transfer, which may be unrealistic. To model travel delay, let \( \tau_{ji} \) be the travel time from \( j \) to \( i \). Then, the number of EVs arriving at \( i \) in time \( t \) should be those dispatched from \( j \) at \( t - \tau_{ji} \):
\[
\sum_{j \in \mathcal{I}} z_{ji}^{t - \tau_{ji}}
\]
The corrected capacity constraint becomes:
\[
0 \leq \beta \cdot \left( \alpha_i^t \cdot F_i^t - \sum_{j \in \mathcal{I}} z_{ij}^t + \sum_{j \in \mathcal{I}} z_{ji}^{t - \tau_{ji}} \right) \leq c_i
\]
This refinement ensures that EVs only contribute to charging demand at the time they actually arrive.
\end{tcolorbox}
  \caption{Dialogues illustrating how the LLM engages with the user across distinct modeling contexts.}
  \label{fig:model-dialogue}
\end{figure*}

\section{Conclusion and future work} \label{Conclusion}
We present an LLM-assisted framework to address the joint investment and assignment problem for EV infrastructure planning, effectively capturing the interdependence and synergy between long-term investment decisions and short-term assignment operations. To overcome computational challenges, we design a distributed ADMM algorithm that enables efficient solution of large-scale, spatial-temporal optimization problems. Our approach is validated using 1.5 million real-world travel records from Chengdu, China. The results demonstrate a 30\% reduction in total system cost compared to a baseline without EV assignment, highlighting the economic benefits of explicitly capturing operational interactions. In addition, the distributed ADMM solver achieves approximately over a 100-fold reduction from minutes to seconds in computational time relative to a centralized solver, making city-scale applications tractable on standard hardware. Together, these findings underline the potential of combining LLM-assisted formulation with distributed optimization to support more effective and scalable EV infrastructure planning.

The proposed model, along with the solution method, holds significant potential for future development. The distributed algorithm facilitates efficient problem-solving through a distributed computing system, where ADMM demonstrates strong adaptability, enabling the model to incorporate a broader range of research factors related to EV charging infrastructure planning. There are several practical challenges warrant careful consideration. For instance, while our empirical analysis uses November data as a baseline month. We acknowledge, however, that demand patterns may vary across months (e.g., due to holidays or weather conditions). Extending the dataset to incorporate multiple months would allow richer capture of seasonal temporal-spatial variation. 
The distributed ADMM framework is expressly designed to scale to extended horizons, enabling future work to capture seasonal temporal-spatial variation when broader datasets are available. Moreover, the charging time for EVs with existing technology is significant, impacting the overall system and user experience \cite{tao_adaptive_2023}. The potential for pile-ups or queues of EVs at charging stations, as well as additional delays arising from traffic congestion or traffic lights, may further increase waiting time and should be accounted for in future research.  Additionally, as time considerations become more intricate, examining the electricity price at varying times and locations is crucial, as they profoundly affect operational costs. Furthermore, although current capabilities remain limited, the rapid development of LLMs suggests the potential for their integration into the design of optimization algorithms in the future, which may lead to further improved efficiency in solving complex problems. In future work, we aim to incorporate a more comprehensive power system that includes power grid, renewable energy sources, and energy storage system. Additionally, we will consider bi-directional charging (e.g. vehicle-to-grid or vehicle-to-vehicle) and charging queue management while integrating stochastic elements into our problem modeling.

\section*{Data availability}
Data will be made available on request.

\section*{Acknowledgement}
This work was supported in part by the Australian Research Council (ARC) Discovery Early Career Researcher Award (DECRA) under Grant DE230100046.

\bibliography{ref.bib}

\end{document}